\documentclass[11pt]{article}
\usepackage{amsxtra,amssymb,amsmath,enumerate,graphicx,bbm,amsthm,color}
\usepackage[active]{srcltx}
\usepackage{t1enc}
\usepackage{aurical}
\usepackage[T1]{fontenc}
\usepackage[]{algorithm2e}

\usepackage{eurosym}

\usepackage{graphicx}
\usepackage{caption}
\usepackage{subcaption}

\newcommand{\e}{\mathbb{E}}
\newcommand{\E}{\mathbb{E}}

\renewcommand{\P}{\mathbb{P}}
\newcommand{\R}{\mathbb{R}}

\newcommand{\1}{{\mathbf 1}}

\def \Esp#1{{\mathbb E}\left[#1\right]}

\def\be{\begin{align}}
\def\ee{\end{align}}
\def\b*{\begin{eqnarray*}}
\def\e*{\end{eqnarray*}}

\def\vp{\varphi}

\makeatletter
\newcommand{\manuallabel}[2]{\def\@currentlabel{#2}\label{#1}}
\makeatother
\newcommand\numberthis{\addtocounter{equation}{1}\tag{\theequation}}

\def\be{\begin{eqnarray}}
\def\ee{\end{eqnarray}}
\def\beq{\begin{equation}}
\def\eeq{\end{equation}}
\def\b*{\begin{eqnarray*}}
\def\e*{\end{eqnarray*}}
\def\bi{\begin{itemize}}
\def\ei{\end{itemize}}

\def \1{{\bf 1}}
\def\vp{\varphi}
\def\eps{\varepsilon}

\def\={\;=\;}

\def\x{\times}

\def\Esp#1{\mathbb{E}\left[#1\right]}

\def \proof{{\noindent \bf Proof. }}
\def \ep{\hbox{ }\hfill$\Box$}

 \def\vs#1{\vspace{#1mm}}

\def \D{\mathbb{D}}
\def \E{\mathbb{E}}
\def \F{\mathbb{F}}

\def \L{\mathbb{L}}
\def \M{\mathbb{M}}
\def \N{\mathbb{N}}
\def \P{\mathbb{P}}

\def \R{\mathbb{R}}

\def \Z{\mathbb{Z}}

\def\Ac{{\cal C}}

\def\Ec{{\cal E}}
\def\Fc{{\cal F}}

\def\Lc{{\cal L}}
\def\Mc{{\cal M}}

\def\Kc{{\cal K}}
\def\Lc{{\cal L}}
\def\Ic{{\cal I}}

\def\Lc{{\cal L}}

\def\Tc{{\cal T}}

\def\Zc{{\cal Z}}

\def\Lb{{\mathbf L}}

\newtheorem{Theorem}{Theorem}[part]

\newtheorem{Proposition}{Proposition}[part]
\newtheorem{Assumption}{Assumption}[part]

\newtheorem{Remark}{Remark}[part]

\makeatletter \@addtoreset{equation}{section}
\def\theequation{\thesection.\arabic{equation}}
\@addtoreset{Definition}{section}

\@addtoreset{Theorem}{section}

\@addtoreset{Proposition}{section}

\@addtoreset{Assumption}{section}

\@addtoreset{Corollary}{section}

\@addtoreset{Lemma}{section}

\@addtoreset{Remark}{section}

\@addtoreset{Example}{section}

\def\L{\xrm}
\def\D{{\rm D}}

\def\w{\hat w}

\def\A{\mathbf{C}}
\def\L{{\mathbf L}}

\begin{document}

\title{Optimal inventory management and order book modeling
\thanks{This research is part of a Cemracs 2017 project and benefited from the support of Initiative de Recherche  ``Strat\'{e}gies de Trading et d'Investissement Quantitatif'',   Kepler-Chevreux and Coll\`{e}ge de France. We are very gratefull for the support of P.~Besson and his team.}}

\author{Nicolas Baradel\footnote{{Universit\'e Paris-Dauphine, PSL Research University, CNRS, CEREMADE, Paris.}  {baradel@ceremade.dauphine.fr}.}
\and Bruno Bouchard\footnote{{Universit\'e Paris-Dauphine, PSL Research University, CNRS, CEREMADE, Paris.}  {bouchard@ceremade.dauphine.fr}. Research of B.~Bouchard partially supported by ANR CAESARS (ANR-15-CE05-
0024). }   \and   {David Evangelista} \footnote{{King Abdullah University of Science and Technology (KAUST), CEMSE
	Division, Thuwal 23955-6900. Saudi Arabia.}  {david.evangelista@kaust.edu.sa}.  D. Evangelista was partially supported by KAUST baseline funds and KAUST OSR-CRG2017-3452.} \and   {Othmane Mounjid} \footnote{CMAP, \'{E}cole Polytechnique. {othmane.mounjid@polytechnique.edu}.}
}
\maketitle

\def\Pa{P^{\mathfrak a}}
\def\Pb{P^{\mathfrak b}}

\def\Ra{R^{\mathfrak a}}
\def\Rb{R^{\mathfrak b}}

\def\Qa{Q^{\mathfrak a}}
\def\Qb{Q^{\mathfrak b}}
\def\sf{S}
\def\Aa{A^{\mathfrak a}}
\def\Ab{A^{\mathfrak b}}
\def\La{L^{\mathfrak a}}
\def\Lb{L^{\mathfrak b}}

\def\Lbtot{{\rm L}^{\mathfrak b}}
\def\Latot{{\rm L}^{\mathfrak a}}

\def\Lai{L^{{\mathfrak a},\frac12}}
\def\Lbi{L^{{\mathfrak b},\frac12}}

\def\Laitot{{\rm L}^{{\mathfrak a},\frac12}}
\def\Lbitot{{\rm L}^{{\mathfrak b},\frac12}}

\def\Aac{\alpha^{\mathfrak a}}
\def\Abc{\alpha^{\mathfrak b}}
\def\Lac{\ell^{\mathfrak a}}
\def\Lbc{\ell^{\mathfrak b}}
\def\Laic{\ell^{{\mathfrak a},\frac12}}
\def\Lbic{\ell^{{\mathfrak b},\frac12}}
\def\M{Z}
\def\Mc{\zeta}
\def\Zbf{{\bf \rm Z}}
\def\Zc{{\mathcal Z}}

\def\Ia{I^{\mathfrak a}}
\def\Ib{I^{\mathfrak b}}
\def\Na{N^{\mathfrak a}}
\def\Nb{N^{\mathfrak b}}
\def\Ba{B^{\mathfrak a}}
\def\Bb{B^{\mathfrak b}}
\def\vs#1{\vspace{#1mm}}
 \def\v{{\rm v}}
 \def\Kc{{\mathcal K}}
 \def\ak{\mathfrak a}
 \def\bk{\mathfrak b}
 
\begin{abstract}
We model the behavior of  three agent classes acting dynamically in a limit order book of a financial asset. Namely, we consider market makers (MM), high-frequency trading (HFT) firms, and institutional brokers (IB). Given a prior dynamic of the order book, similar to the one considered in the Queue-Reactive models \cite{cont2013price,citeulike:12810809,citeulike:13675327}, the MM and the HFT define their trading strategy by optimizing the expected utility of terminal wealth, while  the IB has a prescheduled task to sell or buy many shares of the considered asset. We derive the variational partial differential equations that characterize the value functions of the MM and HFT and explain how almost optimal control can be deduced from them. We then provide a first illustration of the interactions that can take place between these different market participants by simulating   the dynamic of an order book in which each of them plays his own (optimal) strategy. 
\end{abstract}
\maketitle
 
 
\section{Introduction}

The comprehension of the order book dynamic has become a fundamental issue for all market participants and for regulators that try to increase the market transparency and efficiency. A deep understanding of the order book dynamic and agents behaviors enables:  market makers to ensure liquidity provision at  cheaper prices,  high-frequency traders  to reduce arbitrage opportunities, investors to reduce their transaction costs, policy makers to design relevant rules, to strengthen market transparency and to reduce market manipulation. Moreover,  modeling the order book provides insights on the behavior of the price at larger time scales since the price formation process starts at the order book level, see e.g.~\cite{cont2012order} for Brownian diffusion asymptotic of rescaled price processes. Recently, the widespread market electronification has facilitated the access of high quality data describing market participants decisions and interactions at the finest time scale, on which statistics can be based. The availability of the order book data certainly allows a better understanding of the market activity. On the other hand, the recent market fragmentation and the increase of trading frequency rise the complexity of agents actions and interactions. 
  
The main objective of the present paper is to propose a flexible order book dynamics model  close to the  {one of \cite{cont2013price, citeulike:12810809,citeulike:13675327,OptimalBaseLiqOm}},  {construct a first building block towards a realistic order book modeling,} and try to better understand the various regimes related to the presence of different market participants. Instead of considering a pure statistical dynamics as in e.g.~\cite{cont2013price}, we construct a {structural} endogenous dynamics,  { see e.g.~\cite{DiscrOptTrad,OptimalBaseLiqOm}}, based on the optimal behavior of agents that are assumed to be rational. For numerical tractability, we simplify the market in three classes  of  {(most significant)} participants: the market makers (MM), the high-frequency {trading} (HFT) firms, and the institutional brokers (IB).   Each of them decides of his policy in an optimal way, given prior statistics, and   then interact with the others given the endogenous realizations of the market.  

More precisely, we postulate that  they assume an order book model similar to the one suggested by  the Queue-Reactive model \cite{citeulike:12810809}, see also  \cite{abergel2017algorithmic,cont2013price,guilbaud2013optimal}, in which we restrict to the best bid and ask queues\footnote{{This limitation is for numerical tractability. It is already enough for most markets.}}: limit and aggressive orders arrive with certain intensities, when a queue is depleted it is regenerated according to a certain law and possibly after a price move, the spread can take two different values\footnote{One could consider a larger set of possible spreads in theory, but refrains to doing this for numerical tractability.}.  Importantly,  we also take the order book's imbalance into account in the modeling of  the different transition probabilities, see e.g.~Besson et al.~\cite{bes16imb}. 

The   market participants can either put limit or aggressive orders. The aim of a market maker is to gain the spread. He should therefore essentially put limit orders, {aggressive} orders being used when his inventory is too unbalanced. In our model, he can only acts on the given  {order book}. The high-frequency trader is assumed to play on the correlation between the order book dynamics, {viewed as} the stock price, and the price of another asset, called {futures} hereafter. {Indeed, he believes} that the difference between the stock and the futures prices is mean-reverting.  {Whenever he buys/sells one unit of the stock, he sells/buys back one unit of the futures.} {We do not handle} the order book associated to the futures {but simply model} the price of the futures as the mid-price of the stock to which a mean-reverting process is added.  {Still, we introduce a (possibly equal to $0$) transaction cost proportional to the size of the transaction.} As the market-makers, he seeks for a zero inventory at the end of the trading period. Finally, institutional brokers are simply assumed to play VWAP {(Volume Weighted Average Price)}- or Volume-based strategies ({trading algorithms}). Again, they essentially use limit orders and become aggressive when they are too late in their {schedule}. 

\vspace{2mm}

We focus on the derivation of the optimal strategies of the market makers and the high-frequency traders, and on how they can be computed numerically by solving the associated {variational} partial differential equations. Note that it is important to consider their strategies within a dynamical model as current  actions impact the order book and therefore may modify its futures dynamics. We will actually see that, in certain situations, participants can place aggressive orders or limit orders in the spread just to try to manipulate the order book's dynamics in a favorable way.  Note that the dimension of our control space is   higher than in \cite{abergel2017algorithmic,guilbaud2013optimal,DiscrOptTrad,OptimalBaseLiqOm} since more complex decisions are required to tackle the market making problem.

However, the ultimate goal of this work is to provide a market simulator. In the last section, we   already present   simulations of the market behavior given the pre-computed optimal strategies of the different actors. More precisely, we will only simulate the evolution of the mean-reverting process (driving the difference between the stock and the futures price) together with the reconstruction of the queues when prices move,  and let the participants play their optimal strategies given the evolution of the order book {due} to their different actions. This should allow us to study how these different market participants may interact among each other if each of them is playing his optimal policy. In particular, one should observe different market regimes depending on the proportion of the different participants in the total population, on their risk aversion, etc. First simulations are provided in this paper, a more throughout study is left for future research. 

\vspace{2mm}

Our approach therefore lies in between two current streams of literature. The first one is based on ``general equilibrium models'', including economic models, where the market activity is generated by interactions between rational agents who take optimal decisions that interact through the market netting process, see e.g.~\cite{RePEc:eee:finmar:v:2:y:1999:i:2:p:99-134,citeulike:1204538,Rosu2009}. The second stream of literature considers purely statistical models where the order book is seen as a random process, see e.g.~\cite{citeulike:10868255,LOBModHawkes,BayerHorstQiu2015,cont2013price,citeulike:7341957, gueant2017optimal,citeulike:12810809,citeulike:13675327,citeulike:12386824,citeulike:12514252,farmer03a}. The statistical models focus on reproducing many salient features of a real market rather than agents behaviors and interactions.   In our approach, we take into account that the agent's behavior are essentially based on statistical approaches, but that they eventually interact with each other.
 
 \vspace{2mm}
 
 We end this introduction with an outline of our paper. In Section \ref{sec: general order book},  we present the general order book dynamics. The  marker maker control problem is studied in details in Section \ref{sec:marketmaker}. {There,} we present the equations satisfied by {their} optimal strategy and propose a numerical solution for this problem, together with numerical illustrations. In Section \ref{sec: HFT},  we formulate the high-frequency trader control problem and perform a similar analysis. The institutional broker strategy is described in  Section \ref{sec:InstitutionalBroker}, {where} we restrict to VWAP and Volume liquidation problems.  Finally, in Section \ref{sec : Simulation of the global market}, we simulate a realistic market using the three agent's optimal trading strategies.

 
\section{General order book presentation and priors of the market participants}\label{sec: general order book}

As mentioned above, we focus on a single order book and only model the best bid and ask prices, in a similar way to \cite{cont2013price}. In this section, we describe the general market mechanisms as well as the priors on which the optimal strategies of the agents are based. We fix a terminal time horizon $T$ and consider a  probability space $(\Omega,\P)$. Here, $\Omega:=\Omega_{1}\x \Omega_{2}$ and $\P:=\P_{1}\otimes \P_{2}$,  where $\Omega_{1}$ is the space of $\R^{11}$-valued   cadlag paths on $[0,T]$  endowed with a probability measure $\P_{1}$ with full support on $\Omega_{1}$, and $\Omega_{2}$ is the one dimensional Wiener space endowed with the Wiener measure $\P_{2}$. 
\vs2

We denote respectively by $(\Pb_{t})_{t \geq 0}$ and $(\Pa_{t})_{t  \geq 0}$ the best bid offer and the best ask offer processes on the market. They are valued in ${\mathfrak d}\mathbb{Z}$ where ${\mathfrak d} > 0$ is the tick size.  We denote by $(\Qb_{t})_{t \geq 0}$ and $(\Qa_{t})_{t  \geq 0}$ the sizes of the corresponding queues valued in $\mathbb{N}^{*}$. To simplify the notation, we introduce $P := (\Pb, \Pa)$, $Q := (\Qb, \Qa)$ and define the spread process as  $\delta P := \Pa - \Pb$. Moreover, we assume\footnote{The extension to more possible spread values is straightforward. We stick to this setting for notational and computational simplicity. Note that this limit is also justified by empirical evidences for many stocks, see \cite{cont2013price}.} that  $\delta P_{t} \in \{{\mathfrak d},2{\mathfrak d}\}$ for all $t\geq 0$.

\vs2
We denote by $(\tau_{i})_{i\ge 1}$ the times at which orders are {sent} to the market. We assume that this sequence is increasing and that $\#\{i\in \N: \tau_{i}<T\}<\infty$ a.s.
{The market participants} can send different types of orders at each time $\tau_{i}$: 
	\begin{itemize}
		\item {Aggressive} orders of size $\alpha^{\mathfrak b}_{i} \in \mathbb{N} \cap [0, \Qb]$ at the bid or of size $\alpha^{\mathfrak a}_{i} \in \mathbb{N} \cap [0, \Qa]$ at the ask: the size of the corresponding queue, $\Qb$ or $\Qa$, decreases by the size of the aggressive order, $\alpha^{\mathfrak b}_{i}$ or $\alpha^{\mathfrak a}_{i}$.   
				\item Limit orders of size $L^{\mathfrak b}_{i} \in \mathbb{N}$ at the bid or of size $L^{\mathfrak a}_{i}\in \mathbb{N}$ at the ask: the size of the corresponding queue, $\Qb$ or $\Qa$, increases by the size of the limit order, $L^{\mathfrak b}_{i}$ or $L^{\mathfrak a}_{i}$.
		\item When $\delta P = 2{\mathfrak d}$: Limit orders of size $L^{\mathfrak b, \frac{1}{2}}_{i} \in \mathbb{N}$ at the bid or of size $L^{\mathfrak a, \frac{1}{2}}_{i}  \in \mathbb{N}$ at the ask: the order is placed inside the spread, at the price $\Pb + {\mathfrak d} = \Pa - {\mathfrak d}$, this generates a new queue at the bid or at the ask, of size $L^{\mathfrak b, \frac{1}{2}}_{i}$ or $L^{\mathfrak a, \frac{1}{2}}_{i}$, and a price move.
		\item Cancellations: Cancellations of $M^{\bk}_{i} \in \mathbb{N} \cap [0, \Qb]$ orders at the bid or of $M^{\ak}_{i} \in \mathbb{N} \cap [0, \Qa]$ orders at the ask.  {The difference between cancellations and {aggressive} orders is that {aggressive} orders consume the bottom of the limit while we see cancellations as only consuming the top of the limit first.} 
	\end{itemize}

We assume that the sequence    $(\alpha^{\mathfrak b}_{i} ,\alpha^{\mathfrak a}_{i},L^{\mathfrak b}_{i}, L^{\mathfrak a}_{i},L^{\mathfrak b, \frac{1}{2}}_{i},L^{\mathfrak a, \frac{1}{2}}_{i},M^{\bk}_{i},M^{\ak}_{i})_{i\ge 1}$ is made of random variables leaving, with probability one, on the state space $\A_{\circ}$ defined as the collection of elements $(a^{\mathfrak b}, a^{\mathfrak a},\ell^{\mathfrak b},\ell^{\mathfrak a},\ell^{\mathfrak b, \frac{1}{2}},$ $\ell^{\mathfrak a,\frac{1}{2}},m^{\bk},$ $m^{\ak})$ $\in \N^{8}$ such that  
\begin{align}\label{eq: def constraint A}
&\left\{\begin{array}{l}
a^{\mathfrak b} a^{\mathfrak a}=0\\
\ell^{\mathfrak b}= \ell^{\mathfrak a}=0\;\mbox{ if } \max\{a^{\mathfrak b} ,a^{\mathfrak a}\}\ge 1\\
\ell^{\mathfrak b, \frac{1}{2}}=0   \;\mbox{ if } \max\{	a^{\mathfrak b},a^{\mathfrak a},\ell^{\mathfrak b}\}\ge 1 \\
 \ell^{\mathfrak a, \frac{1}{2}}=0 \;\mbox{if } \max\{	a^{\mathfrak b},a^{\mathfrak a},\ell^{\mathfrak a}\}\ge 1 \\
 m^{\mathfrak b }=0   \;\mbox{ if } \max\{	a^{\mathfrak b},a^{\mathfrak a},\ell^{\mathfrak b},\ell^{\mathfrak b,\frac12}\}\ge 1 \\
 m^{\mathfrak a }=0 \;\mbox{if } \max\{	a^{\mathfrak b},a^{\mathfrak a},\ell^{\mathfrak a},\ell^{\mathfrak a,\frac12}\}\ge 1 
\end{array}\right. .
\end{align}
 We interpret the above expression as follows. First, aggressive orders can not be sent simultaneously at the bid and at the ask. Next,  limit orders (at the current bid/ask prices) can not be placed at the same time that aggressive orders are sent. Finally,   one can not place limit orders within the spread if limit orders at the current bid/ask prices are placed. Because we only consider the first limits, these conditions are natural whenever one presumes that orders of different market participants do not arrive exactly at the same time.

Depending on the arrival of orders, queues can be depleted. In this case, new queues can be re-generated, at the same prices or at different prices, and possibly with a change of the spread value. To model this, we introduces a sequence of random variables $(\epsilon_i,\epsilon^{\mathfrak b}_{i},\epsilon^{\mathfrak a}_{i})_{i\ge 1}$ with values in $\{0,1\}\times (\N^{*})^{2}$. The sequence $(\epsilon_i)_{i\ge 1}$ will describe possible jumps of the bid/ask prices, while the sequence   $(\epsilon^{\mathfrak b}_{i},\epsilon^{\mathfrak a}_{i})_{i\ge 1}$ will describe the new sizes of the  queues when they are  re-generated, after one of them is depleted. More precisely, we  postulate the dynamics  
	\begin{align}\label{eq: dyna P Q}
	\begin{array}{rl}
	\Pb_{\tau_i} =& \Pb_{\tau_{i-1}} + {\mathfrak d}\left[\mathbf{1}_{\{\epsilon_i = 1  \}}\left( - \mathbf{1}_{\{\Qb_{\tau_{i-1}} = \hat   \alpha^{\mathfrak b}_{i}  \}} + \mathbf{1}_{\{\delta P_{\tau_{i-1}} =  2{\mathfrak d}\}}\mathbf{1}_{\{\Qa_{\tau_{i-1}} = \hat \alpha^{\mathfrak a}_{i}\}}\right) + \mathbf{1}_{\{L^{\mathfrak b, \frac{1}{2}}_{i} > 0\}}\right] \\
			\Pa_{\tau_i} =& \Pa_{\tau_{i-1}} + {\mathfrak d}\left[\mathbf{1}_{\{\epsilon_i = 1\}}\left(\mathbf{1}_{\{\Qa_{\tau_{i-1}} = \hat \alpha^{\mathfrak a}_{i} \}} - \mathbf{1}_{\{\delta P_{\tau_{i-1}} = 2{\mathfrak d}\}}\mathbf{1}_{\{\Qb_{\tau_{i-1}} = \hat \alpha^{\mathfrak b}_{i}\}}\right) - \mathbf{1}_{\{L^{\mathfrak a, \frac{1}{2}}_{i} > 0\}}\right] \\
			\Qb_{\tau_i} =& \Qb_{\tau_{i-1}} + L^{\mathfrak b}_{i} +  (L^{\mathfrak b, \frac{1}{2}}_{i}-\Qb_{\tau_{i-1}})\mathbf{1}_{\{L^{\mathfrak b, \frac{1}{2}}_{i} > 0\}}-\hat \alpha^{\bk}_{i} \mathbf{1}_{\{\Delta \Pb_{\tau_{i}} = 0\}} +( \epsilon^{\mathfrak b}_{i}-\Qb_{\tau_{i-1}})\mathbf{1}_{\{\Delta \Pb_{\tau_{i}} \ne 0\}\cup\{{\hat \alpha^{\bk}_{i}=\Qb_{\tau_{i-1}}}\}} \\
			\Qa_{\tau_i} =& \Qa_{\tau_{i-1}} + L^{\mathfrak a}_{i} +  (L^{\mathfrak a, \frac{1}{2}}_{i}-\Qa_{\tau_{i-1}})\mathbf{1}_{\{L^{\mathfrak a, \frac{1}{2}}_{i} > 0\}}-\hat \alpha^{\ak}_{i} \mathbf{1}_{\{\Delta \Pa_{\tau_{i}} = 0\}} +( \epsilon^{\mathfrak a}_{i}-\Qa_{\tau_{i-1}})\mathbf{1}_{\{\Delta \Pa_{\tau_{i}} \ne 0\}\cup\{{\hat \alpha^{\ak}_{i}=\Qa_{\tau_{i-1}}}\}}
	\end{array}\end{align}
for $i\ge 1$, where 
$$
\hat \alpha^{\bk/\ak}_{i}:=\alpha^{\bk/\ak}_{i}+M^{\bk/\ak}_{i},
$$
with 
$$
(\Pb_{0},\Pa_{0},\Qb_{0},\Qa_{0})\in D_{P,Q}:=\{(p^{\mathfrak b},p^{\mathfrak a},q)\in ({\mathfrak d}\Z)^{2}\x (\N^{*})^{2}: p^{\mathfrak a}-p^{\mathfrak b}\in \{{\mathfrak d},2{\mathfrak d}\}\},
$$
 and the convention $\tau_{0}=0-$. We refer to \eqref{eq: def constraint A} to see that this dynamics is consistent. In particular, prices can move only if one of the queues is depleted because of the arrival of aggressive orders or if a new limit order is inserted within the spread.  {These two situations can not occur simultaneously}.  The ask price can move by ${\mathfrak d}$ when the ask queue is depleted. If the spread was already $2{\mathfrak d}$, then the bid price moves up as well.  {The other way around if the bid queue is depleted}. In the following, we extend the dynamics of $(P,Q)$ by considering it as a step constant right-continuous process on $[0,T]$.

We now denote by $\Ec$ the   {$\N^{{12}}$-valued} step constant right-continuous process  defined  by 
$$
\Delta \Ec_{\tau_{i}}:=\Ec_{\tau_{i}}-\Ec_{\tau_{i-1}}:=(\alpha^{\mathfrak b}_{i}, \alpha^{\mathfrak a}_{i}, L^{\mathfrak b}_{i}, L^{\mathfrak a}_{i}, L^{\mathfrak b, \frac{1}{2}}_{i}, L^{\mathfrak a, \frac{1}{2}}_{i},{M^{\mathfrak b}_{i}, M^{\mathfrak a}_{i},} \epsilon_i, \epsilon^{\mathfrak b}_{i},\epsilon^{\mathfrak a}_{i},1),\;i\ge 1,
$$
with $\Ec_{\tau_{0}}:=\Ec_{0}:=0$. Later on, we shall only write 
\begin{align}\label{eq: function dyna PQ}
(P_{\tau_{i}},Q_{\tau_{i}})=\Tc_{P,Q}(P_{\tau_{i-1}},Q_{\tau_{i-1}},\Delta \Ec_{\tau_{i}})
\end{align}
in which the map $\Tc_{P,Q}$ is defined explicitly by \eqref{eq: dyna P Q}. 

The process $\Ec$ models the flow of all the orders on the market. From the viewpoint of a market participant, it corresponds to its own orders and to the other participants' orders that we denote by $\tilde \Ec$. The process $\tilde \Ec$ has jump times $(\tilde \tau_{i})_{i\ge 1}\subset (\tau_{i})_{i\ge 1}$ of sizes 
\begin{align}\label{eq: def tilde taui}
\Delta \tilde \Ec_{\tilde \tau_{i}}=\sum_{j\ge 1} \1_{\{\tilde \tau_{i}=\tau_{j}\}}\Delta   \Ec_{ \tau_{j}},\;i\ge 1.
\end{align}
It induces a counting measure $\tilde \nu(dt,de)$. For a market participant, a prior on this {measure} is given by the compensator $\tilde \mu$ of $\tilde \nu$ under $\P$. We assume that it is state dependent. More precisely, we consider a Borel kernel $(p,q)\in D_{P,Q}\mapsto \tilde \mu(\cdot|p,q) \in {\cal M}([0,T]\times \N^{{8}})$, where   ${\cal M}([0,T]\times \N^{{8}})$ denotes the collection of non-negative measures on   $[0,T]\times \N^{{8}}$. In particular, it can depend on the order book's imbalance, as observed in e.g.~\cite{bes16imb}.  To be consistent with the constraints imposed above, it satisfies: 
\begin{align}\label{eq: hyp mu supported by A}
\tilde \mu(\cdot|p,q) \mbox{ is supported by } \A_{\circ}, \;\mbox{ for all } (p,q)\in D_{P,Q}.
\end{align}
It should also be such that aggressive orders  {and cancellations} are never bigger than the corresponding queue size, which will be made more explicit in our numerical example sections, see  Section \ref{subsec: numerics MM}. 

Next, we shall denote by $\Ec^{\phi}$  the flows corresponding to the trading strategy of either a market maker, a high-frequency trader, or an institutional broker.  
Thus,  she will assume facing a global flow $\Ec= \tilde \Ec+ \Ec^{\phi}$. 

Moreover, for simplicity,   we shall  assume that $\tilde \mu$ is of the form
 \begin{align}\label{eq: hyp decompo mu}
d \tilde \mu(c,\eps,\eps^{\mathfrak b},\eps^{\mathfrak a},dt|p,q)=  d\lambda(\eps,\eps^{\mathfrak b},\eps^{\mathfrak a}|p,q,c) d\beta(c|p,q) dt
 \end{align}
 in which $\lambda$ and $\beta$ are bounded Borel non-negative kernels and (without loss of generality)
 \begin{align}\label{eq: normalization beta}
 \int d\lambda(\eps,\eps^{\mathfrak b},\eps^{\mathfrak a}|p,q,c) =1,\;\mbox{ for all } (p,q,c)\in D_{P,Q}\x \A_{\circ}.
 \end{align}
 {Later on, when an order $(\alpha^{\mathfrak b}_{i}, \alpha^{\mathfrak a}_{i}, L^{\mathfrak b}_{i}, L^{\mathfrak a}_{i}, L^{\mathfrak b, \frac{1}{2}}_{i}, L^{\mathfrak a, \frac{1}{2}}_{i},{M^{\mathfrak b}_{i}, M^{\mathfrak a}_{i})}$ is sent by a MM, an HFT or an IB, we shall also assume that the conditional law of $(\epsilon_i, \epsilon^{\mathfrak b}_{i},\epsilon^{\mathfrak a}_{i})$ is given by $\lambda$.}
 
\begin{Remark}\label{rem : number of jumps in small time} Let $\gamma(p,q):=\int d\beta(c|p,q)$. Then,  $\gamma$ is uniformly bounded by the above assumption.  Let $\tau$ be a stopping time and fix $h>0$.  Since $\lambda$ integrates to one, it follows that $\P[\#\{t\in [\tau,\tau+h]: \Delta \tilde \Ec_{t}\ne 0\}=1 |\Fc_{\tau}]=h \gamma(P_{\tau},Q_{\tau})+o(h)$ and $\P[\#\{t\in [\tau,\tau+h]: \Delta \tilde \Ec_{t}\ne 0\}>1 |\Fc_{\tau}]=o(h)$. Moreover, the process counting the number of jumps of $\tilde \Ec$ is dominated by a Poisson process with intensity $\bar \gamma:=\sup \gamma<\infty$. Hence,     $\P[\#\{t\le T: \Delta \tilde \Ec_{t}\ne 0\}\ge k]\le \bar \gamma T/k$, for $k\ge 1$, by Markov's inequality. Similarly, if $g$ is a non-decreasing Borel map, then $\E[g(\#\{t\in [0,T]: \Delta \tilde \Ec_{t}\ne 0\})]\le \sum_{k\ge 1} g(k) \frac{(\bar \gamma T)^{k}}{k!}e^{-\bar \gamma T}$. 
\end{Remark}
 

\section{Market maker's optimal control problem}\label{sec:marketmaker} 

In this section, we describe the optimal control problem of the market maker, the key tools to characterize the solution and how to numerically approximate the optimal control.

\subsection{Market maker's strategy and state dynamics}\label{subsec: strategy MM}

The market maker typically places limit orders in order to make profit of the spread but can turn aggressive when his inventory is too important. At the end the trading period $[0,T]$, the later should be zero. In the following, we denote by $G$ {his} gain process and by $I$ {his} inventory. We also need to keep track of  the sizes  of  {his} orders already placed at the bid queue, $N^{\mathfrak b}$, and at the ask queue, $N^{\mathfrak a}$. For simplicity, we impose that new orders can not be placed at the bid (respectively at the ask) if {he} already has previously taken a position at the {bid} (respectively at the ask). Then, {his} position at the bid (resp.~at the ask) is completely described by  $N^{\mathfrak b}$ (resp.~$N^{\mathfrak a}$) and the number of units $B^{\mathfrak b}$ before {him} in the bid-queue (resp.~$B^{\mathfrak a}$ before {him} in the ask-queue).  Later on, we only write $N=(N^{\mathfrak b},N^{\mathfrak a})$ and $B=(B^{\mathfrak b},B^{\mathfrak a})$.

To define the market maker's control, we assume that {he} faces the exogenous process $\tilde \Ec$ described in Section \ref{sec: general order book}.

 For {him}, a control is a sequence of random variables $\phi=(\tau^{\phi}_{i}, \mathbf{c}^{\phi}_{i})_{i \geq 1}$ where $(\tau^{\phi}_{i})_{i\ge 1}$ is an increasing sequence of  times and each $\mathbf{c}^{\phi}_{i} = (\alpha_{i}^{\mathfrak{b},\phi}, \alpha_{i}^{\mathfrak{a},\phi}, L_{i}^{\mathfrak{b},\phi}, L_{i}^{\mathfrak{a},\phi}, L_{i}^{\mathfrak{b}, \frac{1}{2},\phi}, L_{i}^{\mathfrak{a}, \frac{1}{2},\phi}, M_{i}^{\bk,\phi}, M_{i}^{\ak,{\phi}})$ is  {$\A_{\circ}$-valued}, see below for more implicit restrictions. 
 The times $(\tau_{i}^{\phi})_{i\ge 1}$ are the times at which he sends orders: the action done at $\tau^{\phi}_{i}$ is $\mathbf{c}^{\phi}_{i}$, whose components have the same meanings as in Section \ref{sec: general order book}.  
 
 Given his own orders and the other participants' orders, the sequence of times at which orders are sent to the market is $(\tau_{i})_{i\ge 1}$ where 
 $\tau_{0}:={0-}$ and $\tau_{i+1}=\min\{\tilde \tau_{j}> \tau_{i},\;j\ge 1\}\wedge \min\{ \tau^{\phi}_{j}> \tau_{i},\;j\ge 1\}$, {see \eqref{eq: def tilde taui} and above.}

 We denote by $\Ec^{\phi}$ the {c\`{a}dl\`{a}g} process that  jumps only at the times $\tau_{i}^{\phi}$s with jump size 
 $$
 \Delta \Ec^{\phi}_{\tau^{\phi}_{i}}=(\mathbf{c}^{\phi}_{i}, \sum_{j\ge 1} (\epsilon_j, \epsilon^{\mathfrak b}_{j},\epsilon^{\mathfrak a}_{j})\1_{\{\tau^{\phi}_{i}=\tau_{j}\}}), \;i\ge 1,
 $$ 
 so that\footnote{We keep in mind that $\Ec$ also depends of $\phi$ but dot not make this explicit for ease of notations. } 
 $$
 \Ec=\Ec^{\phi}+\tilde \Ec
 $$
 from his point of view. As usual, we impose that $\Ec^{\phi}$ is   predictable  for the (completed) filtration $\F^{\phi}=(\Fc^{\phi}_{t})_{t\ge 0}$ generated\footnote{{Note that this creates a dependence of the filtration on the control itself, which is similar to \cite{baradel2016optimal}.}} by $\Ec$. We will also keep in mind the number of actions
  $$
 J:=\sum_{i\ge 1} \1_{\{\tau^{\phi}_{i}\le \cdot\}}  
 $$
  from time $0$ on, as it may induce a cost.

 We now impose a minimum and maximum inventory size, denoted by $(-I^{*}, I^{*}) \in (-\N)\times \N$, and that  
$$
\#\{\tau^{\phi}_{i}\le T,i\ge 1\}\le k_{\phi}\wedge J_{\circ}~a.s., \mbox{ for some } k_{\phi}\in \N, 
$$
for some $J_{\circ}\in \N\cup \{\infty\}$. The constraint on the inventory is classic. The constraint on the number of operations can be justified by operational constraints. In the case $J_{\circ}=\infty$, it just means that each control should be of essentially bounded activity, but the bound is not uniform on the set of controls and can be as large as needed.  
 
 To be admissible, a control $\phi$  should therefore be such that each $\mathbf{c}^{\phi}_{i}$ is $\A(Z_{\tau^{\phi}_{i}-})$-valued,
 where 
$$
Z:=(P,Q,X)\;\mbox{ with }\;  X:=(G,I,N,B,J), 
$$ 
and, for $z=(p^{\mathfrak b},p^{\mathfrak a},q^{\mathfrak b},q^{\mathfrak a},g,i,n^{\mathfrak b},n^{\mathfrak a},b^{\mathfrak b},b^{\mathfrak a},j)$, $\A(z)$ is the collection of elements $c:=(a^{\mathfrak b}, a^{\mathfrak a},\ell^{\mathfrak b},$ $\ell^{\mathfrak a},\ell^{\mathfrak b, \frac{1}{2}},$ $\ell^{\mathfrak a, \frac{1}{2}},$ $m^{\bk},$ $m^{\ak})\in \A_{\circ}\setminus\{0\}$ such that :
 	\begin{align*}
		a^{\mathfrak{b}} \leq \min\left\{ i +I^{*} - b^{\mathfrak{a}}  \;;\; q^{\mathfrak b} \right\} \;,&\;
		   a^{\mathfrak{a}} \leq \min\left\{ I^{*} - i - b^{\mathfrak{b}}   \;;\; q^{\mathfrak a}\right\},\\
		 \ell^{\mathfrak{b}} \leq ( I^{*} -  i)\1_{\{{n^{\bk}= 0}\}}   \;,&\;\ell^{\mathfrak{a}} \leq( i +I^{*})\1_{\{  n^{\ak}= 0\}} , \\
		 	\ell^{\mathfrak{b}, \frac12} \leq  (I^{*} - i)\1_{\{p^{\ak}-p^{\bk}=2{\mathfrak d}\}} \1_{\{{n^{\bk}= 0}\}}\;,&\;   
		   \ell^{\mathfrak{a}, \frac12} \leq  ( i +I^{*})\1_{\{p^{\ak}-p^{\bk}=2{\mathfrak d}\}}\1_{\{{n^{\ak}= 0}\}} \\
		   m^{\bk}\le n^{\mathfrak b}\;&,\;m^{\ak}\le n^{\mathfrak a}\\
		   c=0 &\mbox{ if } j=J_{\circ}.
	\end{align*}
	Note that the constraints on the {first three lines} correspond  to the fact that we do not want to take a position that could lead to an inventory out of the limits $-I^{*}$ and $I^{*}$ if it was suddenly executed. The indicator functions correspond to   additional constraints on the controls, imposed for numerical tractability: no new limit order can be sent on a side if one has not been executed or  has not cancelled the position on the same side before, no limit order can be sent in the spread if it is not equal to two ticks. 
	 In this case, we write $\phi\in \Ac(0,Z_{0-})$.

The dynamics of $X$ is given by 
 \begin{align}\label{eq: dyna X}
 X_{\tau_{i}}=\Tc_{X}(P_{\tau_{i-1}},Q_{\tau_{i-1}},X_{\tau_{i-1}},\Delta \Ec^{\phi}_{\tau_{i}})\1_{\{\Delta \Ec^{\phi}_{\tau_{i}}\ne 0\}}+\tilde \Tc_{X}(P_{\tau_{i-1}},Q_{\tau_{i-1}},X_{\tau_{i-1}},\Delta \tilde \Ec_{\tau_{i}})\1_{\{\Delta \Ec^{\phi}_{\tau_{i}}= 0\}},
 \end{align}
in which {$\Tc_{X},\tilde \Tc_{X}: \R^{22}\mapsto {\mathfrak d} \Z\times \N\times \N^{4}\times\N$}. More precisely, consider the map
$$
 {{\rm exe}(a,n,b):=\min\{(a - b)^{+},n\}},\;a,b,n\in \N.
 $$
It represents the number of stocks set at a limit that are executed when an aggressive order of size $a$ arrives, that the position in the queue is $b$, and the size of the posted block at this position is $n$. 

Then, having in mind the constraints encoded in $\A(\cdot)$ above, see also \eqref{eq: def constraint A}, we can write  
$$
\Tc_{X}=(\Tc_{G},\Tc_{I},\Tc_{N^{\bk}},\Tc_{N^{\ak}},\Tc_{B^{\bk}},\Tc_{B^{\ak}},\Tc_{J})\;,\;\tilde\Tc_{X}=(\tilde\Tc_{G},\tilde\Tc_{I},\tilde\Tc_{N^{\bk}},\tilde\Tc_{N^{\ak}},\tilde\Tc_{B^{\bk}},\tilde\Tc_{B^{\ak}},\tilde\Tc_{J})
$$
where  
 \begin{align*}
  &\Tc_{G}(p,q,x,\delta)= g+(a^{\mathfrak b}-{\rm exe}(a^{\mathfrak b},n^{\bk},b^{\bk}))p^{\mathfrak b}-(a^{\mathfrak a}-{\rm exe}(a^{\mathfrak a},n^{\ak},b^{\ak}))p^{\mathfrak a} \\
 &\Tc_{I}(p,q,x,\delta)= i-(a^{\mathfrak b}-{\rm exe}(a^{\mathfrak b},n^{\bk},b^{\bk}))+(a^{\mathfrak a}-{\rm exe}(a^{\mathfrak a},n^{\ak},b^{\ak}))\\
& \Tc_{N^{\bk/\ak}}(p,q,x,\delta)= n^{\bk/\ak}+[\ell^{\bk/\ak}-n^{\bk/\ak}]^{+} +\ell^{\bk/\ak,\frac12} -m^{\bk/\ak}   - {\rm exe}(a^{\bk/\ak},n^{\bk/\ak},b^{\bk/\ak}) \\
 &\Tc_{B^{\bk/\ak}}(p,q,x,\delta)=  b^{\bk/\ak}+   (q^{\bk/\ak}-b^{\bk/\ak}) \1_{\{\ell^{\bk/\ak}\ne 0\}}-b^{\bk/\ak}\1_{\{m^{\bk/\ak}=n^{\bk/\ak}\}}-(b^{\bk/\ak}\wedge a^{\bk/\ak})\1_{\{a^{\bk/\ak}\ne 0\}}-b^{\bk/\ak}\1_{\{\ell^{\bk/\ak,\frac12}\ne 0\}}\\
&\Tc_{J}(p,q,x,\delta) =j+1
 \end{align*}
 and 
  \begin{align*}
& \tilde \Tc_{G} (p,q,x,\delta)=g-  {\rm exe}(a^{\bk},n^{\bk},b^{\bk})p^{\mathfrak b} + {\rm exe}(a^{\ak},n^{\ak},b^{\ak})p^{\mathfrak a} \\
& \tilde \Tc_{I}(p,q,x,\delta)=i+  {\rm exe}(a^{\bk},n^{\bk},b^{\bk}) - {\rm exe}(a^{\ak},n^{\ak},b^{\ak})\\
&  \tilde \Tc_{N^{\bk/\ak}}(p,q,x,\delta)= n^{\bk/\ak}-   {\rm exe}(a^{\bk/\ak},n^{\bk/\ak},b^{\bk/\ak})\\
 &   \tilde \Tc_{B^{\bk/\ak}}(p,q,x,\delta)=[b^{\bk/\ak}-a^{\bk/\ak}]^{+}\1_{\{m^{\bk/\ak}= 0\}}+(b^{\bk/\ak}-[m^{\bk/\ak}-(q^{\bk/\ak}-b^{\bk/\ak}-n^{\bk/\ak})]^{+})^{+}\1_{\{m^{\bk/\ak}\ne 0\}}\\
 &  \tilde \Tc_{J}(p,q,x,\delta)=0,
 \end{align*}
  for $x=(g,i,n^{\mathfrak b},n^{\mathfrak a},b^{\mathfrak b},b^{\mathfrak a},j)$, $\delta=(a^{\mathfrak b}, a^{\mathfrak a},\ell^{\mathfrak b},\ell^{\mathfrak a},\ell^{\mathfrak b, \frac{1}{2}},\ell^{\mathfrak a, \frac{1}{2}},m^{\bk},m^{\ak},\eps,\eps^{\mathfrak b},\eps^{\mathfrak a})$, $p=(p^{\mathfrak b},p^{\mathfrak a})$ and $q=(q^{\mathfrak b},q^{\mathfrak a})$
. 
 
%

 \begin{Remark} It follows from \eqref{eq: hyp decompo mu} and the constraint that $\Ec^{\phi}$ is predictable that the probability that $\Ec^{\phi}$ and $\tilde \Ec$ jump at the same time on $[0,T]$ is zero. 
 This justifies the formulation \eqref{eq: dyna X}.
 \end{Remark}

For later use, note that it follows from  \eqref{eq: dyna P Q} and \eqref{eq: dyna X} that 
\begin{align}\label{eq: dyna P Q X}
 Z_{\tau_{i}}=\Tc(Z_{\tau_{i-1}},\Delta \Ec^{\phi}_{\tau_{i}})\1_{\{\Delta \Ec^{\phi}_{\tau_{i}}\ne 0\}}+\tilde \Tc(Z_{\tau_{i-1}},\Delta \tilde \Ec_{\tau_{i}})\1_{\{\Delta \Ec^{\phi}_{\tau_{i}}= 0\}},
\end{align}
in which 
\begin{align*}
\Tc=(\Tc_{P,Q},\Tc_{X})\; \mbox{ and } \;\tilde \Tc=(\Tc_{P,Q},\tilde \Tc_{X}). 
\end{align*}

  \subsection{The optimal control problem}\label{subsec: optimal control MM}

The aim of the market maker is to maximize her expected utility 
$$
\E[U(Z_{T}) ]
$$
in which  
\begin{equation}\label{eq: utility MM}
U(z):=-\exp\left(-\eta\{ g +i^{+} p^{\mathfrak b}- i^{-}p^{\mathfrak a}   - \kappa ( [i^{+}-q^{\mathfrak b}]^{+}+[i^{-}-q^{\mathfrak a}]^{+})- \varrho j \} \right)
\end{equation}
for $z=(p^{\mathfrak b},p^{\mathfrak a},q^{\mathfrak b},q^{\mathfrak a},g,i,n^{\mathfrak b},n^{\mathfrak a},b^{\mathfrak b},b^{\mathfrak a},j)$. In the above, $\eta>0$ is the absolute risk aversion parameter, and $\kappa>0$ is a penalty term taking into account that liquidating the current inventory may lead to a worse price than the one corresponding to the best bid or ask: the quantity $i^{+} p^{\mathfrak b}- i^{-}p^{\mathfrak a}$ corresponds to the liquidation value of the inventory if the bid and ask queues are big enough to absorb it, the expression starting from $\kappa$   takes  into account the number of shares that will not be liquidated at the best limit. The coefficient $\varrho\ge 0$ penalizes the number of actions taken by the market maker. 

To define the corresponding value function, we now extend the definition of our state processes by writing 
$$
Z^{t,z,\phi}=(P^{t,z,\phi},Q^{t,z,\phi},X^{t,z,\phi})
$$ for the process satisfying \eqref{eq: dyna P Q}-\eqref{eq: dyna X} for the control $\phi$ and with initial condition $Z^{t,z,\phi}_{t-}=z\in D_{Z}$ where 
$
D_{Z}$ is the collection of elements $(p^{\mathfrak b},p^{\mathfrak a},q^{\mathfrak b},q^{\mathfrak a},g,i,n^{\mathfrak b},n^{\mathfrak a},b^{\mathfrak b},b^{\mathfrak a},j)\in \D_{P,Q}\x {\mathfrak d} \Z  \times \{-I^{*},\ldots, I^{*}\}\times \N^{4}\x \{0,\ldots,J_{\circ}\}$ such that
\begin{align*}
n^{\bk}+i\le I^{*}\;&,\;i-n^{\ak}\ge -I^{*}\\
b^{\bk}+n^{\bk}\le q^{\bk}\;&,\;b^{\ak}+n^{\ak}\le q^{\ak}.
\end{align*} 

The corresponding  set of admissible controls is $\Ac(t, z)$, and the filtration associated to $\phi\in \Ac(t, z)$  is $\F^{t,z,\phi}$. 
We then set 
$$
\v(t,z):=\sup_{\phi \in \Ac(t, z)}J(t,z;\phi)\;\mbox{ for } (t,z)\in [0,T]\x D_{Z},
$$
where 
$$
J(t,z;\phi):=\E[U(Z^{t,z,\phi}_{T})].
$$

\begin{Remark}\label{rem: dependence on g}
For later use, observe that 
$$
\v(t,z)={e^{-\eta g}} \v(t,p^{\mathfrak b},p^{\mathfrak a},q^{\mathfrak b},q^{\mathfrak a},0,i,n^{\mathfrak b},n^{\mathfrak a},b^{\mathfrak b},b^{\mathfrak a},{j})
$$
for all $t\le T$ and  $z=(p^{\mathfrak b},p^{\mathfrak a},q^{\mathfrak b},q^{\mathfrak a},g,i,n^{\mathfrak b},n^{\mathfrak a},b^{\mathfrak b},b^{\mathfrak a},j)\in D_{Z}$. {Moreover, if $J_{\circ} = \infty$, we also have
$$
\v(t,z)=e^{-\eta (g-\varrho j)} \bar \v(t,z):=e^{-\eta (g-\varrho j)} \v(t,p^{\mathfrak b},p^{\mathfrak a},q^{\mathfrak b},q^{\mathfrak a},0,i,n^{\mathfrak b},n^{\mathfrak a},b^{\mathfrak b},b^{\mathfrak a},0)
$$
}
\end{Remark}

\begin{Remark}\label{rem: v bounded} Note that $\v$ is bounded from above by $0$ by definition. On the other hand, for all $z=(p^{\mathfrak b},p^{\mathfrak a},q^{\mathfrak b},q^{\mathfrak a},g,i,n^{\mathfrak b},n^{\mathfrak a},b^{\mathfrak b},b^{\mathfrak a},j)\in D_{Z}$,
\begin{align*}
\v(t,z)&\ge \min_{i\in [{-I^{*}},I^{*}]} \E[U(P^{t,z,0}_{T},0,0,g,i,0,0,0,0,j)]\\
&= e^{-\eta( g-\varrho j)} \min_{i\in [-I^{*},I^{*}]} \E[U(P^{t,z,0}_{T},0,0,0,i,0,0,0,0,0)],
\end{align*}
where $P^{t,z,0}$ corresponds to the dynamics in the case that the MM does not act on the order book up to $T$. Moreover, it follows from \eqref{eq: utility MM} that 
\begin{align*}
\E[U(P^{t,z,0}_{T},0,0,0,i,0,0,0,0,0)]&\ge -e^{\eta I^{*} |p^{\bk}|} \E[e^{\eta I^{*} (|P^{t,z,0,\bk}_{T}-p^{\bk}|+2{\mathfrak d}+\kappa) }]
\end{align*}
where 
$$
\sup_{p^{\bk}\in {\mathfrak d}\Z}\E[e^{\eta I^{*} (|P^{t,z,0,\bk}_{T}-p^{\bk}|+2{\mathfrak d}+\kappa) }]<\infty, 
$$
by   Remark \ref{rem : number of jumps in small time} and the fact that the price can jump only by ${\mathfrak d}$ when a market event occurs. {Thus,} $\v$ belongs to the class $\L_{\infty}^{{\rm exp}}$ of functions $\vp$ such that $\vp/{\rm L}$ is bounded, in which 
$$
{\rm L}(p^{\bk},g{,j}):=e^{-\eta (g-I^{*}|p^{\bk}|-{\varrho j})}
$$ 
for $(p^{\bk},g,{j})\in {\mathfrak d}\Z \x{\mathfrak d}\Z\x {\{0,\ldots,J_{\circ}\}}$.
\end{Remark}
 
 \subsection{The dynamic programming equation}\label{subsec: DPP equation MM}
 
The derivation of the dynamic programming equation is standard, and is based on the dynamic programming principle. We state below the weak  version of Bouchard and Touzi \cite{bouchard2011weak}, we let $\v_{*}$ and $\v^{*}$ denote the lower- and upper-semicontinuous envelopes of $\v$. 

\begin{Proposition}\label{prop: DPP MM}
Fix $(t,z)\in [0,T]\x D_{Z}$ and a family $\{\theta^{\phi}, \phi\in \Ac(t, z)\}$ such that each $\theta^{\phi}$ is a  $[t,T]$-valued $\F^{t,z,\phi}$-stopping time and $\|Z^{t,z,\phi}_{\theta^{\phi}}\|_{{\mathbf L}_{\infty}}<\infty$. Then, 
\begin{align*}
 \sup_{\phi \in \Ac(t, z)} \mathbb{E}\left[\v_{*}(\theta^{\phi}, Z^{t,z,\phi}_{\theta^{\phi}})\right]
\leq  \v(t, z) \leq \sup_{\phi \in \Ac(t, z)} \mathbb{E}\left[\v^{*}(\theta^{\phi}, Z^{t,z,\phi}_{\theta^{\phi}}) \right].
\end{align*}
\end{Proposition}

\proof The right-hand side inequality follows from a conditioning argument, see  \cite{bouchard2011weak}. The left-hand side is more delicate because the set of admissible controls depends on the initial data. However, it can be easily proved along the lines of  Baradel et al.~{\cite{baradel2016optimal}} when $\{\theta^{\phi}, \phi\in \Ac(t, z)\}$ is $[t,T]\cap (\N\cup \{t,T\})$-valued. Then, the general case is obtained by approximating $[t,T]$-valued stopping times from the right (recall that $Z$ is right-continuous). 
\ep 
\vs2

One can then derive the corresponding dynamic programming equation.  For $z=(p,q,x)\in D_{Z}$,  $c\in \A(z)$, $t\le T$, and a continuous and bounded function $\vp$, we set 
$$
\Ic\vp(t,z):=\int (\Kc^{c}\vp(t,z)-\vp(t,z))d\beta(c|p,q) \mbox{ and } \Kc\vp(t,z):=\sup_{c\in \A(z)}\Kc^{c}\vp(t,z)
$$
where we use the convention that $\Kc^{0}=\sup\{\emptyset\} =-\infty$, and 
$$
\Kc^{c}\vp(t,z):=\int \vp(t,\Tc(z,c,\eps,\eps^{\mathfrak b},\eps^{\mathfrak a}))d\lambda(\eps,\eps^{\mathfrak b},\eps^{\mathfrak a}|p,q,c),
$$
recall \eqref{eq: hyp decompo mu}, \eqref{eq: normalization beta} and \eqref{eq: dyna P Q X}. 

 \vs2

The partial differential equation characterization of $\v$ is then at least formally given by 
 \begin{equation}\label{eq: edp MM}
    \begin{aligned}
    					\min\left\{-\partial_{t}\varphi -\Ic\varphi, \varphi -\Kc\varphi\right\} &= 0 \text{ on } [0, T) \times D_{Z} \\
                        \min\left\{\varphi -U, \varphi -\Kc\varphi\right\}&= 0 \text{ on } \{T\} \times D_{Z}.
     \end{aligned}
    \end{equation}

 In order to ensure that the above is correct, we need two additional conditions. 
\begin{Assumption}\label{ass: thm visco MM} For all upper-semicontinuous (resp.~lower-semicontinuous)    $\vp\in \L_{\infty}^{{\rm exp}}$, the map $(t,z)\in [0,T]\x D_{Z}\mapsto ( \Ic,\Kc)\varphi(t,z)$ is  upper-semicontinuous (resp.~lower-semicontinuous) and belongs to $\L_{\infty}^{{\rm exp}}$.
\end{Assumption}

\begin{Assumption}\label{ass: comp MM} There exists a Borel  function  $\psi$ that is continuously differentiable in time and such that 
\begin{enumerate}[\rm (i)]
\item $0\ge \partial_{t}\psi +   \Ic\psi$ on $[0,T)\x D_{Z}$, 
\item $\psi-  \Kc\psi\ge \iota$ on   $[0,T]\x D_{Z}$ for some $\iota>0$,
\item $\psi\ge   U$ on $\{T\}\x D_{Z}$,
\item $\liminf\limits_{n\to \infty} (\psi/{\rm L})(t_{n},z_{n})=\infty$ if  $|z_{n}|\to \infty$ as $n\to \infty$, for all $(t_{n},z_{n})_{n\ge 1}\subset [0,T]\x D_{Z}$.
\end{enumerate}
\end{Assumption}

Then, one can actually prove that $\v$ is the unique solution of \eqref{eq: edp MM} in the class $\L_{\infty}^{\rm exp}$ defined in Remark \ref{rem: v bounded}.

\begin{Theorem}\label{thm : visco MM} Let Assumption \ref{ass: thm visco MM} hold. Then, $\v_{*}$ (resp.~$\v^{*}$) is a viscosity supersolution (resp.~subsolution) of  \eqref{eq: edp MM}. If moreover Assumption \ref{ass: comp MM} holds, then   $\v$ is continuous on $[0,T)\x D_{Z}$ and is the unique viscosity solution of  \eqref{eq: edp MM}, in the class of (discontinuous) solutions in $\L_{\infty}^{{\rm exp}}$.
\end{Theorem}

\proof  In view of  Proposition \ref{prop: DPP MM}, the derivation of the viscosity super- and subsolution properties is very standard under Assumption \ref{ass: thm visco MM}, see e.g.~ \cite{baradel2016optimal,bouchard2011weak}. 
As for uniqueness, let us assume that $v$ and $w$ are respectively a super- and a subsolution.   Let $\psi$ be as in  Assumption {\ref{ass: comp MM}}. Then,    $(v-w-\psi )(t_{n},z_{n})$ converges to $-\infty$ if  $|z_{n}|\to \infty$ as $n\to \infty$, for any sequence $(t_{n},z_{n})_{n\ge 1}\subset [0,T]\x D_{Z}$, 
and showing that $v\ge w$ on $[0,T]\x D_{Z}$ can be done by, e.g., following the line of arguments of   \cite[Proposition 5.1]{baradel2016optimal}. {Finally, $\v^{*},\v_{*}\in \L_{\infty}^{{\rm exp}}$ by Remark  \ref{rem: v bounded}.}
\ep

\begin{Remark}\label{rem : cond suff comp MM} If $J_{\circ}<\infty$ and the supports of $\lambda(\cdot|p,q,c)$ and $\gamma(\cdot|p,q)$ are bounded, uniformly in $(p,q,c)\in ({\mathfrak d}\Z)^{2}\x\N^{2}\x \A_{\circ}$, then it is not difficult to see that the function defined by 
$$
\psi(t,z):=e^{2\eta (1+I^{*})|z|}e^{-r (j+t)}, \mbox{ for } z=(p,q,g,i,n,b,j)\in D_{Z}\mbox{ and } t\le T,
$$
satisfies the requirements of Assumption \ref{ass: comp MM}, for $r$ large enough. Verifying Assumption \ref{ass: comp MM} in the case $J_{\circ}=\infty$ seems much more difficult. On the other hand, the sequence of value functions associated to a sequence $(J_{\circ}^{n})_{n\ge 1}$ increases to the value function associated to $J_{\circ}=\infty$ as $J_{\circ}^{n}\to  \infty$.  which provides a natural way to construct a convergent numerical scheme for the computation of $\v$ and the optimal control policy, see Sections \ref{subsec: numerics MM} and \ref{subsec: approximate control MM} below. {Standard arguments based on this approximation would also imply that $\v_{*}=\v$ and that $\v$ is the smallest supersolution of  \eqref{eq: edp MM}, in the class of (discontinuous) solutions in $\L_{\infty}^{{\rm exp}}$.}
\end{Remark}
\subsection{Dimension reduction, symmetries and numerical resolution}\label{subsec: numerics MM}

Before to provide a converging numerical scheme for \eqref{eq: edp MM}, let us first  recall that the variables $g$ {(}and $j$ {if $J_{\circ} = \infty$)} can be omitted, see Remark \ref{rem: dependence on g}.
 If moreover, the transition kernels depend   on prices only through the spread (which is a natural assumption at least on a rather short time horizon), then one more dimension can be eliminated.
 
 \begin{Assumption}\label{ass: dep kernel by spread MM}
The  kernel $(p,q,c)\in D_{P,Q}\x \A \mapsto (\lambda(\cdot|p,q,c),\beta(\cdot|p,q))$ depend on $p=(p^{\mathfrak b},p^{\mathfrak a})$ only through the value of the mid-spread $\delta p:=(p^{\mathfrak a}-p^{\mathfrak b})/2$.
  \end{Assumption}
Indeed, if Assumption \ref{ass: dep kernel by spread MM} holds, then one easily checks that 
\begin{align*}
e^{-\eta i p^{\circ}} \bar \v(t,-\delta p,\delta p,q^{\mathfrak b},q^{\mathfrak a},0,i,\cdot)&=\bar \v(t,p^{\circ}-\delta p,p^{\circ}+\delta p,q^{\mathfrak b},q^{\mathfrak a},0,i,\cdot)\\
&=\bar \v(t,p^{\mathfrak b},p^{\mathfrak a},q^{\mathfrak b},q^{\mathfrak a},0,i,\cdot)
\end{align*}
with $p^{\circ}:=(p^{\mathfrak a}+p^{\mathfrak b})/2$, so that 
$$
e^{\eta i (p^{\mathfrak b}+\delta p)}\bar \v(t,p^{\mathfrak b},p^{\mathfrak b}+2\delta p,q^{\mathfrak b},q^{\mathfrak a},0,i,\cdot)
$$
does not depend on $ p^{\mathfrak b}$ but only on $\delta p$.  

The resolution of the equation can also be simplified by using potential symmetries, in the sense of  the following assumption.
\begin{Assumption}\label{ass: symmetrie MM} For all $(p,q)\in D_{P,Q}$, 
$c:=(a^{\mathfrak b}, a^{\mathfrak a},\ell^{\mathfrak b},\ell^{\mathfrak a},\ell^{\mathfrak b, \frac{1}{2}},$ $\ell^{\mathfrak a, \frac{1}{2}},m^{\mathfrak b},m^{\mathfrak a})\in \A$ and all Borel sets $O\subset \{0,1\}$, $O^{\mathfrak b},O^{\mathfrak a}\subset \N$, 
\begin{align*}
\int_{O\x O^{\mathfrak b}\x O^{\mathfrak a}} d\lambda(\eps,\eps^{\mathfrak b},\eps^{\mathfrak a}|p,q,c)=\int_{O\x O^{\mathfrak a}\x O^{\mathfrak b}} d\lambda(\eps,\eps^{\mathfrak a},\eps^{\mathfrak b}|\bar p,\bar q,\bar c)
\end{align*}
where $\bar p=(-p^{\ak},-p^{\bk})$, $\bar q=(q^{\mathfrak a},q^{\mathfrak b})$ and $\bar c=(a^{\mathfrak a},a^{\mathfrak b}, \ell^{\mathfrak a},\ell^{\mathfrak b},\ell^{\mathfrak a, \frac{1}{2}},\ell^{\mathfrak b, \frac{1}{2}},m^{\mathfrak a},m^{\mathfrak b})$. Moreover,  for all Borel sets 
$O=O^{\mathfrak b}_{a}\x O^{\mathfrak a}_{a}\x O^{\mathfrak b}_{\ell}\x O^{\mathfrak a}_{\ell}\x O^{\mathfrak b}_{\ell^{\frac12}}\x O^{\mathfrak a}_{\ell^{\frac12}}\x O^{\mathfrak b}_{m}\x O^{\mathfrak a}_{m}\subset \N^{8}$, 
\begin{align*}
\int_{O}d\beta(c|p,q) =\int_{\bar O}d\beta(c|\bar p,\bar q)
\end{align*}
where   $\bar O:=O^{\mathfrak a}_{a}\x O^{\mathfrak b}_{a}\x O^{\mathfrak a}_{\ell}\x O^{\mathfrak b}_{\ell}\x O^{\mathfrak a}_{\ell^{\frac12}}\x O^{\mathfrak b}_{\ell^{\frac12}}\x O^{\mathfrak a}_{m}\x O^{\mathfrak b}_{m}$. 
\end{Assumption}
The above assumption implies that the transition probabilities of the order book are symmetric at the bid and at the ask, whenever the configurations are. Then, $\v$ admits a symmetry which can be exploited to reduce the complexity of the numerical resolution of \eqref{eq: edp MM}. Namely,  under Assumption  \ref{ass: symmetrie MM}, we have 
\begin{align*}
  \bar \v(t,p^{\bk},p^{\ak},q^{\mathfrak b},q^{\mathfrak a},0,i,n^{\mathfrak b},n^{\mathfrak a},b^{\mathfrak b},b^{\mathfrak a},j)&=  \bar \v(t,-p^{\ak},-p^{\bk},q^{\mathfrak a},q^{\mathfrak b},0,-i,n^{\mathfrak a},n^{\mathfrak b},b^{\mathfrak a},b^{\mathfrak b},j). 
\end{align*}

Let us now turn to the definition of a numerical scheme for \eqref{eq: edp MM}. We now make the additional assumption that the supports of $\lambda$ and $\beta$ are bounded (not that they are already discrete, by nature).
\begin{Assumption}\label{ass : finite support} $J_{\circ}<\infty$ and there exists  finite Borel sets $O_{1}\subset \N^{8}$ and $O_{2}\subset \N^{3}$ such that $\beta(\cdot|p,q)$ is supported by $O_{1}$ and $\lambda(\cdot|p,q,c)$ is supported by  $O_{2}$, for all $(p,q,c)\in D_{P,Q}\x \A$.   
\end{Assumption}
Then, the operators $\Ic$ and $\Kc$ are explicit. Hence, the only required discretization is in time. For a time step $T/n>0$, we define a time grid $\pi_{n}:=\{t^{n}_{i},i\le n\}$ where 
 $t^{n}_{i}=iT/n$ for $i\le n$. We next consider the sequence of space domains $D_{Z}^k:=D_{Z}\cap [-k,k]^{11}$ for $k\ge 1$, and we let $\v^{k}_{n}$ be the solution of 
  \begin{equation}\label{eq: edp MM num scheme}
    \begin{aligned}
    			&\v^{k}_{n}(t^{n}_{i},\cdot)=\max\left\{\v^{k}_{n}(t^{n}_{i+1},\cdot)+\frac{T}{n}\Ic \v^{k}_{n}(t^{n}_{i+1},\cdot), \max_{c\in \A(\cdot)} \Kc^{c,n} \v^{k}_{n}(t^{n}_{i+1},\cdot)  \right\} = 0 \mbox{ on }   D_{Z}^k,\;i\le n-1 \\
                        &\v^{k}_{n} -U  = 0 \text{ on } (\{T\} \times D_{Z}^k)\cup (\pi_{n}\times  (D_{Z}\setminus D_{Z}^k)),
     \end{aligned}
    \end{equation}
    where 
    $$
    \Kc^{c,n}=\Kc^{c}+\frac{T}{n}\Ic\circ\Kc^{c}. 
    $$
This fully explicit scheme is convergent.
\begin{Proposition}\label{prop: conv vnk MM} Let Assumption \ref{ass : finite support} hold, then the sequence $(\v^{k}_{n})_{k,n\ge 1}$ converges pointwise  to $\v$ on $[0,T)\x D_{Z}$ as $k,n\to \infty$.
\end{Proposition}

\proof First note that $(\v^{k}_{n}/{\rm L})_{k,n\ge 1}$ is uniformly bounded, where ${\rm L}$ is defined in Remark \ref{rem: v bounded}. This follows from Assumption \ref{ass : finite support} and a simple induction argument, compare with Remark \ref{rem: v bounded}. Then, standard stability results, see e.g.~\cite{barles1991convergence} and \cite[Section 3.2]{baradel2016optimalMML}, imply that the relaxed upper-limit $\v^{\infty}$ and lower-limit of $\v_{\infty}$ of $(\v^{k}_{n})_{k,n\ge 1}$ are respectively sub- and supersolution of \eqref{eq: edp MM} and belong to the class $\L^{\rm exp}_{\infty}$. The comparison result mentioned in the proof of Theorem \ref{thm : visco MM}, see Remark \ref{rem : cond suff comp MM}, thus implies that $\v_{\infty}\ge \v\ge \v^{\infty}$ while $ \v_{\infty}\le \v^{\infty}$ by definition. \ep 

\subsection{Approximate optimal controls}\label{subsec: approximate control MM}

In the following, we estimate the optimal control in a  classical way. 
For each $i<n$, we choose a measurable map $\hat c_{n}^{k}(t^{n}_{i},\cdot)$  such that 
\begin{align*}
&\hat c_{n}^{k}(t^{n}_{i},\cdot)\in {\rm arg}\max \{\Kc^{c,n} \v^{k}_{n}(t^{h}_{i+1},\cdot),\;c\in \A(\cdot)\}\;\mbox{ on } D_{Z}^k\\
&\hat c_{n}^{k}(t^{n}_{i},\cdot)=0 \mbox{ on } D_{Z}\setminus D_{Z}^k,
\end{align*}
and define the sequence of stopping times 
$$
\hat \tau^{n,k}_{j+1}:=\min\{t^{n}_{i}: i\ge 0, t^{n}_{i}>\hat \tau^{n,k}_{j},(\v^{k}_{n}-\Kc^{\hat c_{n}^{k}}\v^{k}_{n}(t^{n}_{i+1},\cdot)) (t^{n}_{i},\hat Z^{n,k}_{t^{n}_{i}-})=0\},\;j\ge 0,
$$
with $\hat \tau^{n,k}_{0}:=0-$, and in which $\hat Z^{n,k}=(\hat P^{n,k},\hat Q^{n,k},\hat X^{n,k})$ is defined as in \eqref{eq: dyna P Q X} for the initial condition $Z_{0-}$ at $0$ and the control associated to $\hat \phi^{k}_{n}:=(\hat \tau^{n,k}_{i},\hat c_{n}^{k}(\hat \tau^{n,k}_{i},\hat Z^{n,k}_{\hat \tau^{n,k}_{i}-}))_{i\ge 1}$ in a Markovian way.  
\vs2
This provides a sequence of controls that is asymptotically optimal. 

\begin{Proposition}\label{prop : conv control opti MM} Let the conditions of Proposition \ref{prop: conv vnk MM} hold. Then,
$$
\lim_{n,k\to \infty} \E[U(\hat Z^{n,k}_{T})]=\v(0,Z_{0-}). 
$$
\end{Proposition}

\proof 
Let $\gamma(p,q):=\int d\beta(c|p,q)$ and recall that $\gamma$ is uniformly bounded by assumption, as well as the sequence $(\v^{k}_{n})_{k,n\ge 1}$.  Then, it follows from Remark \ref{rem : number of jumps in small time}    that 
\begin{align*}
\v^{k}_{n}(t^{n}_{i+1},\hat Z^{n,k}_{t^{n}_{i}})+\frac{T}{n}\Ic \v^{k}_{n}(t^{n}_{i+1},\hat Z^{n,k}_{t^{n}_{i}})
=&\v^{k}_{n}(t^{n}_{i+1},\hat Z^{n,k}_{t^{n}_{i}})(1-\frac{T}{n} \gamma(\hat P^{n,k}_{t^{n}_{i}},\hat Q^{n,k}_{t^{n}_{i}}))\\
&+ \frac{T}{n} \int \Kc^{c} \v^{k}_{n}(t^{n}_{i+1},\hat Z^{n,k}_{t^{n}_{i}})d\beta(c|\hat P^{n,k}_{t^{n}_{i}},\hat Q^{n,k}_{t^{n}_{i}})\\
=&\E[\v^{k}_{n}(t^{n}_{i+1},\tilde \Tc(\hat Z^{n,k}_{t^{n}_{i}},\Delta \tilde \Ec_{t^{n}_{i+1}}))|\Fc^{\hat \phi^{k}_{n}}_{t^{n}_{i}}]
+o(n^{-1} )
\end{align*}
and{, similarly,} 
\begin{align*}
\Kc^{\hat c_{n}^{k}(\hat \tau^{n,k}_{i},\hat Z^{n,k}_{\hat \tau^{n,k}_{i}-}),n}\v^{k}_{n}(\hat \tau^{n,k}_{i}+\frac{T}{n},\hat Z^{n,k}_{\hat \tau^{n,k}_{i}-})
=&\E[\v^{k}_{n}(\hat \tau^{n,k}_{i}+\frac{T}{n},\tilde \Tc(\hat Z^{n,k}_{\hat \tau^{n,k}_{i}+\frac{T}{n}-},\Delta \tilde \Ec_{\hat \tau^{n,k}_{i}+\frac{T}{n}}))|\Fc^{\hat \phi^{k}_{n}}_{\hat \tau^{n,k}_{i}-}]+o(n^{-1})
\end{align*}
with the convention that $\tilde \Tc(\cdot,0)$ is the identity. Let $\theta^{n}_{k}$ be the first time when $\hat Z^{n,k}$ exists the domain $D_{Z}^{k}$. In view of \eqref{eq: edp MM num scheme},  Assumption \ref{ass : finite support} {(in particular that $J_{0}<\infty$)} and  Remark \ref{rem : number of jumps in small time} {(see the arguments at the end of Remark \ref{rem: v bounded})}, an induction implies that 
$$
\v^{k}_{n}(0,Z_{0-})=\E[U(\hat Z^{n,k}_{\theta^{n}_{k}})]+o_{n}(1)=\E[U(\hat Z^{n,k}_{T})]+o_{n}(1)+o_{k}(1)
$$
in which $o_{n}(1)$ and $o_{k}(1)$ go to $0$ as $n\to \infty$ and $k\to \infty$.   It remains to appeal to Proposition \ref{prop: conv vnk MM}. \ep

\subsection{Numerical experiments}\label{subsec: num expe MM}
 
We now turn to a numerical experiment.  We compute an approximation of the  optimal control as described in Section \ref{subsec: approximate control MM}, using the simplifications detailed in Section \ref{subsec: numerics MM}. 
\vs2

Let us first describe a realistic prior distribution for the evolution of $\tilde \Ec$. The coefficients we use are inspired   from  the behavior of the stock Soci\'et\'e G\'en\'erale (CLE FP)\footnote{We thank Chevreux-Kepler for providing us these data.} and from \cite{bes16imb}.   

As for the prior on the dynamics of the market. We simply consider that both market and limit orders  arrive according to a Poisson process.  Both limit and aggressive orders arrive with an intensity of $0.6$  per second. When a limit order arrives, it is assigned to the bid or the ask with probability ${1/2}$. When a market order arrives, we assign it to the bid or the ask according to the statistics described for big caps in \cite[Chart 8 p.10]{bes16imb}. Namely, if 
$$
{\rm Imb_{\tilde \tau_{i}}}:=\frac{ \Qb_{\tilde \tau_{i}}-\Qa_{\tilde \tau_{i}} }{\Qb_{\tilde \tau_{i}}+\Qa_{\tilde \tau_{i}} }
$$
is the order book imbalance at the time ${\tilde \tau_{i}}$ at which the market order arrives, then it arrives at the ask with probability {$0.5+0.35*{\rm Imb}_{\tilde \tau_{i}}$}.

To describe the size of the orders and of the inventory, we take $1/2$ of the ATS ({mediAn}\footnote{{Following \cite[Page 23]{bes16imb}}.} Trade Size)   as the unit.

The size of the market orders is also assumed to be dependent on the order book imbalance. We use  \cite[Chart 16 p.10]{bes16imb} to estimate that the size of the trade arriving at the ask (if it arrives at the ask), represents a percentage  of the queue (that we round to an integer number, by above). Namely, we set {$\hat {\rm f}^{\ak}_{\tilde \tau_{i}}:=0.7+0.3*{\rm Imb}_{\tilde \tau_{i}}$}, and assign a (conditional) probability of {$60\%$} that the order is of size  ${\rm round}[\hat {\rm f}^{\ak}_{\tilde \tau_{i}}*\Qa_{\tilde \tau_{i}}]$ and the same probability that the executed volume deviates from the latter by one unit (with equal probability to be by one more and one less unit - if quantities are negative or bigger than the size of the queue, we obviously set them to $0$ or to the size of the queue).

We use a simpler {modeling approach} for the limit orders. With {$55\%$} probability a limit order (if it arrives) is of size of $2$ (recall that the unit is $1/2$ of an ATS). It is of size $3$ with probability {$10\%$} and of size $1$ with probability {$35\%$}.      {Again, it is based on \cite{bes16imb}. }

{When a queue is depleted}, the probability of a price move is set to {$\P[\epsilon_{i}=1|\Fc_{\tilde \tau_{i}}]= {75}\%$}.  If the bid is depleted but the bid price does not go  {down}, the {size of the new bid queue is set to $2$ units with probability $60\%$, $1$ unit with probability $25\%$ and $3$ units with probability $15\%$, otherwise it is set to $10$ units with probability $60\%$, $5$ units with probability $25\%$ and $12$ units with probability $15\%$. The same applies to the ask price if this is the ask queue that is depleted}.

 If both ask and bid prices move, recall \eqref{eq: function dyna PQ}, we take the same distribution for both queues and consider them as being (conditionally) independent. The distribution corresponds to the one of a price move, as described above. 

When the spread is equal to two ticks, the next {limit order arrives in the spread with probability $90\%$}. {This models the fact that a spread of two ticks is not common, see e.g.~\cite{cont2013price}.}  To be consistent with the probability of arrival of market orders at the bid or at the ask, we assume that  {a new bid limit is created in the spread with probability $0.5+0.35*{\rm Imb}_{\tilde \tau_{i}}$} (otherwise, this is a new ask limit). {The size of this new limit is $2$ with  $55\%$,  of size $3$ with probability {$10\%$} and of size $1$ with probability $35\%$. This corresponds to  the behavior of the stock Soci\'et\'e G\'en\'erale (CLE FP).}    {We do not change the dynamic of aggressive orders with the size of the spread.} 

\vs2

Let us now describe the other parameters of the Market Maker's optimal control problem. 
We set the  {bid} price to $10$ at time $0-$, the spread is one tick, and the tick equal to {0.01}, recall from Section \ref{subsec: numerics MM} that only the {spread} size matters {(because we have here the required symmetry)} and observe that the latter can be rescaled together with the level of risk aversion, which is here fixed to {$\eta=1$. We take $\kappa=0.02$ and $\varrho=10^{-20}$}. The time horizon is $T=59$ seconds, and the time step is $1$ second. We keep a small time horizon for a better visibility of the evolution of the order book. We do not consider cancellations from the rest of the market for simplicity. Moreover, in order to reduce the computation time, we add the additional constraint that the MM can not cancel only part of position in a queue, he can only cancel the whole position. We also fix a maximal queue size  of ${12}$  and fix the maximal absolute value of the inventory to $I^{*}={7}$. The size of the limit and market orders sent by the MM are constrained to be less than  {3}. This corresponds to adding an additional constraint in the definition of $\Ac(\cdot)$ which does not change the above analysis.

\vs2

\begin{figure}[h!]
\centering
 \includegraphics[width=0.5\textwidth]{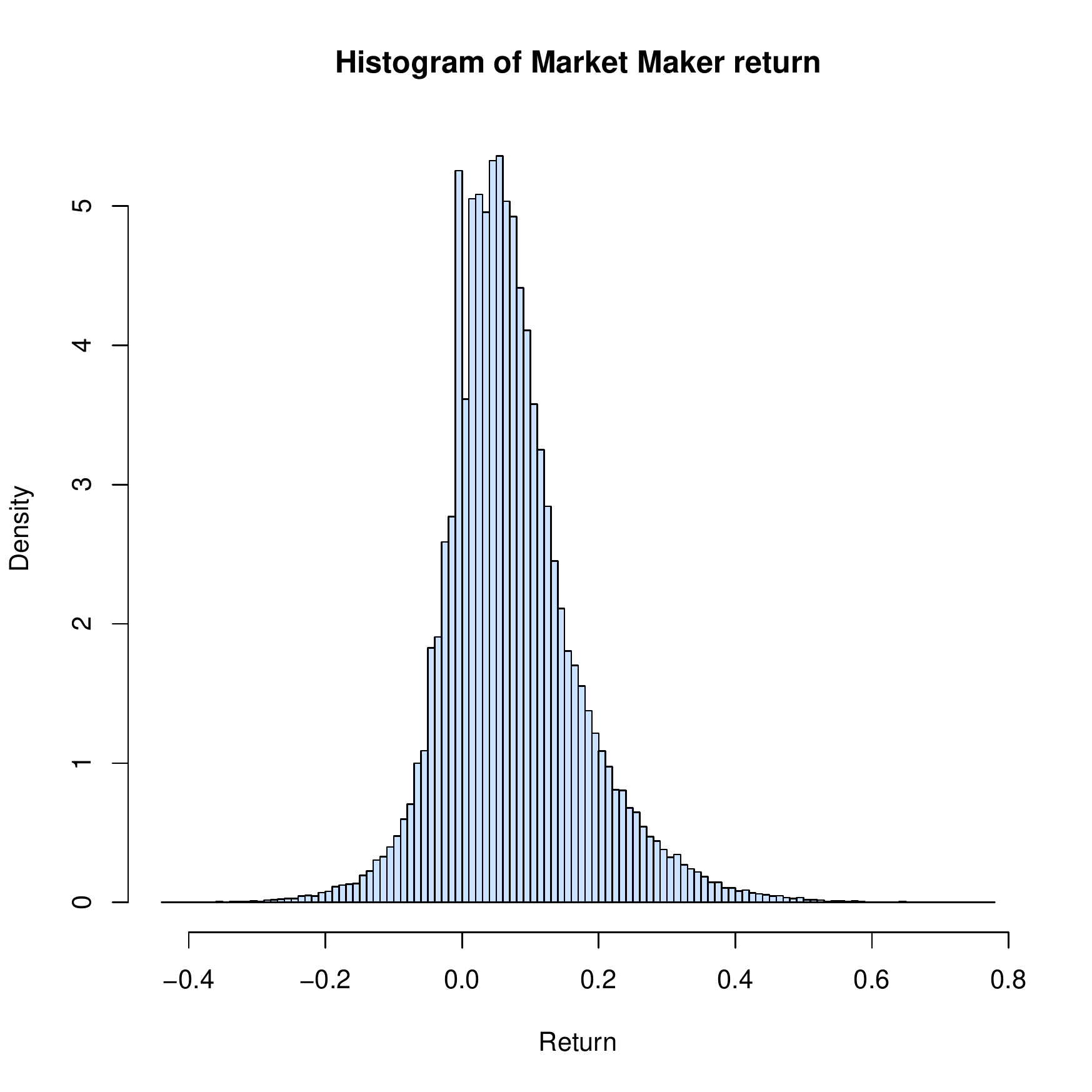}
 \caption{Density estimation of the gain made by  the Market Maker.} 
\label{fig:HistMM}
\end{figure}

\begin{figure}[h!]

 \includegraphics[width=\textwidth,height=0.86\textheight]{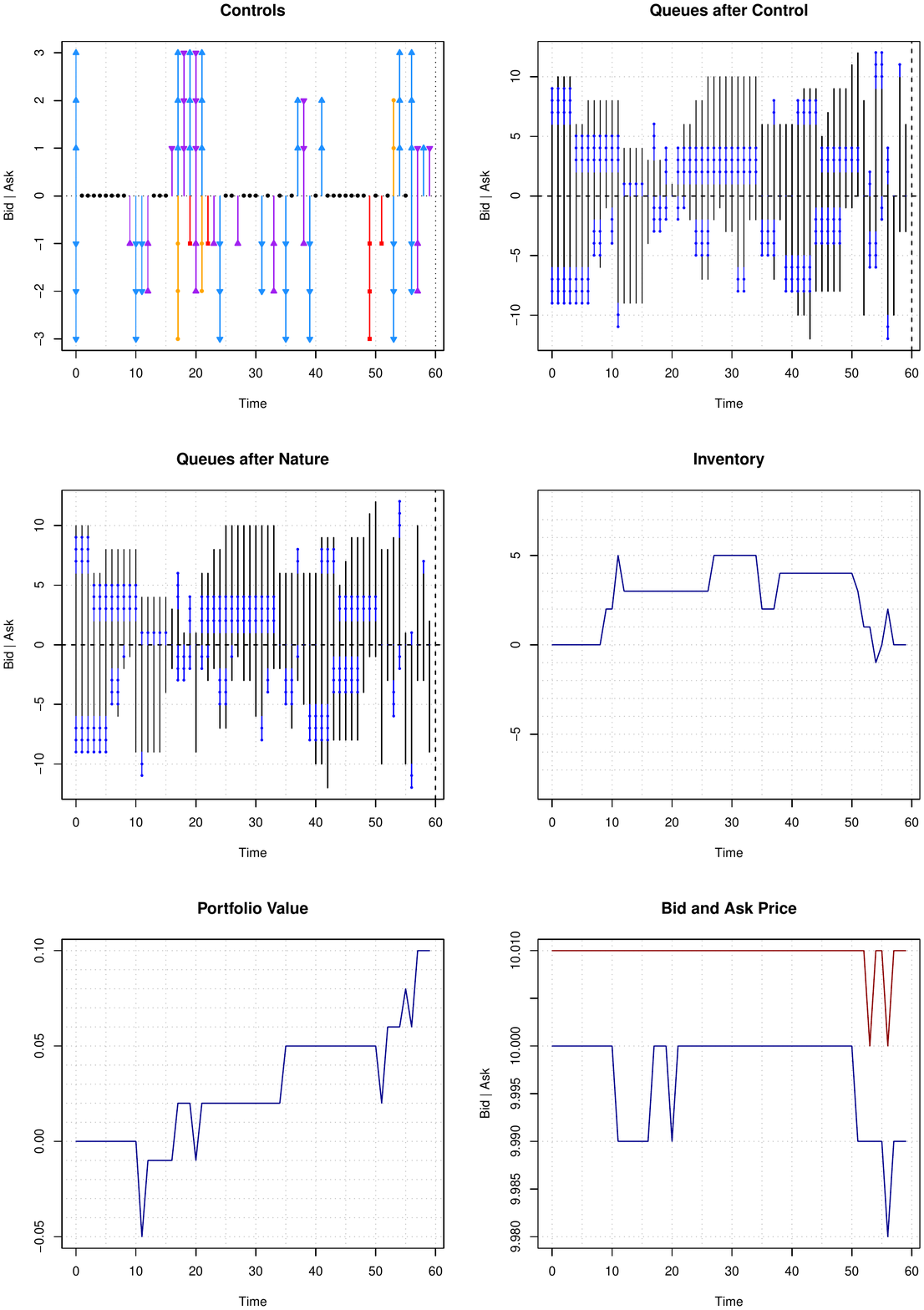}
 \caption{Simulated path of the optimal strategy of the Market Maker.} 
\label{fig:MM75}
\end{figure}

In Figure \ref{fig:MM75}, we provide a simulated path of the optimal strategy, starting from a symmetric configuration of the order book, with queue lengths equal to $6$. In this simulation, the MM always play before the other (random) players\footnote{This is just a convention, since the transition probabilities do not depend on time.}. The top left graphic describes the control played by the MM. Triangles pointing outward (with respect to the zero line) correspond to limit orders, the number of triangles giving their size. Arrows with triangles pointing inward are cancellations, again the number of triangles gives the size. Aggressive orders are symbolized by lines with squares, while limit orders within the spread correspond to the lines with dots. The top right graphic gives the state of the order book just after the MM has played, and before the nature (i.e.~the other players) plays. The size of the lines gives the total length of each queues, while the dots symbolize the position of the MM in the queue. The middle left graphic describes the state of the order book after the nature plays. The bottom left graphic gives the value of the portfolio of the MM if he had to liquidate (by sending aggressive orders) his stock holding at the best bid/ask price. The final value gives the ``true'' liquidation value of the book. In the case that the final inventory can not be liquidated at the best bid/ask price, we liquidate the remaining part at the best bid/ask price minus/plus one tick. The middle right graphic is the inventory just before he plays and the bottom right one is the bid and ask prices just after he plays.  Figure \ref{fig:HistMM} provides the distribution of the gain made by the MM, it uses $10^{5}$ simulated paths.

His strategy can seem  difficult to interpret at first sight. But, we have to remember that he believes that the imbalance plays an important role in the book order dynamics, and that he not only should take care of it but that he can actually use it: some limit orders are sent not to be executed but to influence the evolution of  the price in a favorable direction. Also note that he should avoid the price to go down/up if his inventory is positive/negative. Having this in mind, it is not surprising that he can be sometimes at the limit of a price manipulation strategy, {if he is big enough with respect to the market, which is the case in these simulations}. Finally, we have to keep in mind that his strategy is constrained. He sometimes cancels positions to be free to react more quickly to a more favorable market configuration later on. 
  
Not surprisingly, the MM first sends limit orders of equal sizes on both sides of the order book. Nothing happens until time $t=9$. At this time is inventory is $2$ and his position at the ask is still far from the beginning of the queue. In order to avoid increasing again his inventory, he cancels his remaining  position of one unit at the bid. He puts a new position at the bid at time $10$ after the queue has slightly increased,  because of an exogenous limit order. Unfortunately for him, this new position is immediately executed, and his inventory jumps to $5$. After the queue is regenerated, he immediately puts a limit order of size $2$ at the bid. His reasoning is the following: 1.~this position has little chance to be executed immediately, he does not take a risk of again increasing his inventory; 2.~by doing this he increases the imbalance and therefore the probability of being executed at the ask so has to gain the spread and reduce his inventory. This strategy is successful since immediately $2$ units of his positions at the ask are executed. The imbalance is still good even if he cancels his last unit at the bid, so as to be free to play the control he wants once his last unit at the ask will be executed. In fact, it does not work and he decides to refresh his global position at time $16$ by canceling his final unit at the ask and putting limit orders again at time $17$ in a symmetric way. This is a limit spread order at the bid. This makes sense since he has to avoid the stock price to go down, because he has a positive inventory. The fact that just after he alternates between putting and canceling limit orders at the ask is probably a numerical artifact: on the one hand, he wants to have a limit position at the ask because his inventory is large, on the other hand he does not want to increase the imbalance to avoid having a too big probability of being executed at the bid and of seeing the price go down. Similarly, he wants to keep a position at the bid to avoid a downward price move, while he really does not want to be executed on this side. The aggressive order at the bid at time $19$ is just a partial cancellation, that is completed at time $20$. He just does not want to be executed. By doing so, he unfortunately causes a downward jump of the bid, which is not good for him. Just after he sends a limit spread order at the bid to fight against this downward pressure, and then cancels it once the bid is back at a distance one tick of the ask. This strategy is successful. The rest of his orders can be interpreted in a similar manner. Just note that he starts to be aggressive at the bid side at time $49$ because his inventory is too big and the maturity starts to be quite close, in particular the first market order of size $3$ compensates the execution of a position  of size $3$ at the bid just before (so that the graphic of the inventory does not move, although the inventory temporary jumps to the upper limit $7$).

\newpage
\newpage

 
 \section{High frequency trader's   pair trading problem}\label{sec: HFT}
We now turn to the  HFT strategy. We consider a pair trading strategy where the trader invests in the difference of two highly correlated assets. Here, we choose the futures price of the stock to be the second asset. The order book defined in Section \ref{sec: general order book} represents the dynamics of the stock. Whenever the trader buys/sells $n$ units of stocks (being by an aggressive order or by the execution of a limit order), she sells/buys immediately $n$ units of the futures. Her inventory should be fully liquidated at $T$. 

 \subsection{The optimal control problem}
 
 We assume that the reference price  $F$ of the futures  is given by  
\begin{align}\label{eq: def F}
F=\frac{P^{\mathfrak b}+P^{\mathfrak a}}{2}+S 
\end{align}
 where $S$ is a mean-reverting process
\begin{align}\label{eq: dyna S}
S= S_{0}+\int_{0}^{\cdot}\rho(\hat s-S_t)dt+\int_{0}^{\cdot}\sigma(S_{t}) dW_t.
\end{align}
Here,  $\rho$ is the strength of mean reversion, $\hat s$ is the average of mean reversion and $\sigma:\R\mapsto \R$ is a Lipschitz  bounded function representing the volatility of the process. In the following, we shall assume that 
\begin{align}\label{eq: ass sigma bounded support}
\mbox{ the support of $\sigma$ is bounded.}
\end{align}
\begin{Remark}\label{rem : DS}
The above implies in particular that $S$ lies in a certain compact set $D_{S}$ as soon as $S_{0}$ does. {This could clearly be relaxed to the price of a finer analysis. }
\end{Remark}

The strategy of the HFT is described by the same quantities as the one of the MM in Section \ref{sec:marketmaker}. The only difference lies in the fact that she constantly holds a number equal to $-I$ units of the futures $F$. We assume that buying/selling the futures leads to the payment of a proportional cost $\kappa\ge 0$. Then, the dynamics of the corresponding state process $X$ is given by 
 \begin{align}\label{eq: dyna X HFT}
 X_{\tau_{i}}=\Tc_{X}(S_{\tau_{i}},P_{\tau_{i-1}},Q_{\tau_{i-1}},X_{\tau_{i-1}},\Delta \Ec^{\phi}_{\tau_{i}})\1_{\{\Delta \Ec^{\phi}_{\tau_{i}}\ne 0\}}+\tilde \Tc_{X}(S_{\tau_{i}},P_{\tau_{i-1}},Q_{\tau_{i-1}},X_{\tau_{i-1}},\Delta \tilde \Ec_{\tau_{i}})\1_{\{\Delta \Ec^{\phi}_{\tau_{i}}= 0\}},
 \end{align}
 with $\Tc_{X}$ and $\tilde \Tc_{X}$ now defined with respect to 
  \begin{align*}
  &\Tc_{G}(s,p,q,x,\delta)= g+(a^{\mathfrak b}-{\rm exe}(a^{\mathfrak b},n^{\bk},b^{\bk}))\Delta^{\mathfrak b}_{-}-(a^{\mathfrak a}-{\rm exe}(a^{\mathfrak a},n^{\ak},b^{\ak}))\Delta^{\mathfrak a}_{+} \\
 &\Tc_{I}(s,p,q,x,\delta)= i-(a^{\mathfrak b}-{\rm exe}(a^{\mathfrak b},n^{\bk},b^{\bk}))+(a^{\mathfrak a}-{\rm exe}(a^{\mathfrak a},n^{\ak},b^{\ak}))\\
& \Tc_{N^{\bk/\ak}}(s,p,q,x,\delta)= n^{\bk/\ak}+[\ell^{\bk/\ak}-n^{\bk/\ak}]^{+} +\ell^{\bk/\ak,\frac12} -m^{\bk/\ak}   - {\rm exe}(a^{\bk/\ak},n^{\bk/\ak},b^{\bk/\ak}) \\
 &\Tc_{B^{\bk/\ak}}(p,q,x,\delta)=  b^{\bk/\ak}+   (q^{\bk/\ak}-b^{\bk/\ak}) \1_{\{\ell^{\bk/\ak}\ne 0\}}-b^{\bk/\ak}\1_{\{m^{\bk/\ak}=n^{\bk/\ak}\}}-(b^{\bk/\ak}\wedge a^{\bk/\ak})\1_{\{a^{\bk/\ak}\ne 0\}}-b^{\bk/\ak}\1_{\{\ell^{\bk/\ak}\ne 0\}}\\
&\Tc_{J}(s,p,q,x,\delta) =j+1
 \end{align*}
 and 
  \begin{align*}
& \tilde \Tc_{G} (s,p,q,x,\delta)=g-  {\rm exe}(a^{\bk},n^{\bk},b^{\bk})\Delta^{\mathfrak b}_{+} + {\rm exe}(a^{\ak},n^{\ak},b^{\ak})\Delta^{\mathfrak a}_{-} \\
& \tilde \Tc_{I}(s,p,q,x,\delta)=i+  {\rm exe}(a^{\bk},n^{\bk},b^{\bk}) - {\rm exe}(a^{\ak},n^{\ak},b^{\ak})\\
&  \tilde \Tc_{N^{\bk/\ak}}(s,p,q,x,\delta)= n^{\bk/\ak}-   {\rm exe}(a^{\bk/\ak},n^{\bk/\ak},b^{\bk/\ak})\\
  &   \tilde \Tc_{B^{\bk/\ak}}(p,q,x,\delta)=[b^{\bk/\ak}-a^{\bk/\ak}]^{+}\1_{\{m^{\bk/\ak}= 0\}}+(b^{\bk/\ak}-[m^{\bk/\ak}-(q^{\bk/\ak}-b^{\bk/\ak}-n^{\bk/\ak})]^{+})^{+}\1_{\{m^{\bk/\ak}\ne 0\}}\\
&  \tilde \Tc_{J}(s,p,q,x,\delta)=0,
 \end{align*}
 in which 
 $$
 \Delta^{\mathfrak b}_{\pm}:=p^{\mathfrak b}-\frac{p^{\mathfrak b}+p^{\mathfrak a}}{2}-s
\pm\kappa \mbox{ , }  \Delta^{\mathfrak a}_{\pm}:=p^{\mathfrak a}-\frac{p^{\mathfrak b}+p^{\mathfrak a}}{2}-s
\pm\kappa.
 $$
In the above, we used the notations    $x=(g,i,n^{\mathfrak b},n^{\mathfrak a},b^{\mathfrak b},b^{\mathfrak a},j)$, $\delta=(a^{\mathfrak b}, a^{\mathfrak a},\ell^{\mathfrak b},\ell^{\mathfrak a},\ell^{\mathfrak b, \frac{1}{2}},\ell^{\mathfrak a, \frac{1}{2}},m^{\bk},$ $m^{\ak},$ $\eps,\eps^{\mathfrak b},\eps^{\mathfrak a})$, $p=(p^{\mathfrak b},p^{\mathfrak a})$ and $q=(q^{\mathfrak b},q^{\mathfrak a})$. 

The set of admissible controls $\Ac(0,S_{0},Z_{0-})$ is defined as in Section \ref{sec:marketmaker} but with respect to the (completed) filtration $\F^\phi=(\Fc^{\phi}_{t})_{t\ge 0}$ generated by $(S,\Ec)$.  
\vs2

We also assume that she has an exponential type utility function, with risk aversion parameter $\eta>0$. Then, she wants to maximize over $\phi\in \Ac(0,S_{0},Z_{0-})$ the expected utility 
$$
\E[U(  S_{T},Z_{T}^{\phi})]
$$
where
\begin{equation}\label{utility}
U(s,z):=-\exp\left(-\eta\{ g +i^{+}  \Delta^{\mathfrak b}_{-}- i^{-} { \Delta^{\mathfrak a}_{+} }  - \kappa ( [i^{+}-q^{\mathfrak b}]^{+}+[i^{-}-q^{\mathfrak a}]^{+}) -\varrho j\} \right)
\end{equation}
for $z=(p^{\mathfrak b},p^{\mathfrak a},q^{\mathfrak b},q^{\mathfrak a},g,i,n^{\mathfrak b},n^{\mathfrak a},b^{\mathfrak b},b^{\mathfrak a},j)$, and where $\Delta^{\mathfrak b}_{\pm}$ and $\Delta^{\mathfrak a}_{\pm}$ are defined as above. 
\vs2

As in Section \ref{sec:marketmaker},   we  next extend the definition of our state processes by writing 
$$
(S^{t,s},Z^{t,s,z,\phi})=(S^{t,s},P^{t,s,z,\phi},Q^{t,s,z,\phi},X^{t,s,z,\phi})
$$ for the process satisfying \eqref{eq: dyna P Q}-\eqref{eq: dyna X}-\eqref{eq: dyna S} for the control $\phi$ and the initial condition $(S^{t,s}_{t},Z^{t,s,z,\phi}_{t-})$ $=$ $(s,z)\in D_{S}\x D_{Z}$. The corresponding  set of admissible controls is $\Ac(t, s,z)$, and the filtration associated to $\phi\in \Ac(t, s,z)$  is $\F^{t,s,z,\phi}=(\Fc^{t,s,z,\phi}_{s})_{s\in [t,T]}$. 
We finally define the value function 
$$
\v(t,s,z):=\sup_{\phi \in \Ac(t, s,z)}J(t,s,z;\phi)\;\mbox{ for } (t,s,z)\in [0,T]\x D_{S}\x D_{Z},
$$
where 
$$
J(t,s,z;\phi):=\E[U(S^{t,s}_{T},Z^{t,s,z,\phi}_{T})].
$$

We close this section with Remarks that are the counterparts of Remarks \ref{rem: dependence on g} and \ref{rem: v bounded}.
\begin{Remark}\label{rem: dependence on g HFT}
For later use, observe that 
$$
\v(t,s,z)=e^{-\eta (g-\varrho j)} \bar \v(t,z):=e^{-\eta (g-\varrho j)} \v(t,s,p^{\mathfrak b},p^{\mathfrak a},q^{\mathfrak b},q^{\mathfrak a},0,i,n^{\mathfrak b},n^{\mathfrak a},b^{\mathfrak b},b^{\mathfrak a},0)
$$
for all $t\le T${, $s\in D_{S}$} and  $z=(p^{\mathfrak b},p^{\mathfrak a},q^{\mathfrak b},q^{\mathfrak a},g,i,n^{\mathfrak b},n^{\mathfrak a},b^{\mathfrak b},b^{\mathfrak a},j)\in D_{Z}$. 
\end{Remark}

\begin{Remark}\label{rem: v bounded HFT} Note that $\v$ is bounded from above by $0$ by definition. It   also follows from  \eqref{eq: ass sigma bounded support} that $ S^{t,s} $ takes values in the compact set  $D_{S}$   so that $\v\in \L^{\rm exp}_{\infty}$ by the same reasoning as in  Remark \ref{rem: v bounded}.
\end{Remark}

 \subsection{The dynamic programming equation}
 
As in Section \ref{subsec: DPP equation MM}, we first provide a dynamic programming principle. Again, we let $\v_{*}$ and $\v^{*}$ denote the lower- and upper-semicontinuous envelopes of $\v$.

\begin{Proposition}\label{prop: DPP HFT}
Fix $(t,s,z)\in [0,T]\x D_{S}\x D_{Z}$ and a family $\{\theta^{\phi}, \phi\in \Ac(t, s,z)\}$ such that each $\theta^{\phi}$ is a $[t,T]$-valued $\F^{t,s,z,\phi}$-stopping time and 
$\|(S^{t,s}_{\theta^{\phi}},Z^{t,s,z,\phi}_{\theta^{\phi}})\|_{\infty}<\infty$. Then,  
\begin{align*}
 \sup_{\phi \in \Ac(t, x,q)} \mathbb{E}\left[\v_{*}(\theta^{\phi}, S^{t,s}_{\theta^{\phi}},Z^{t,s,z,\phi}_{\theta^{\phi}})\right]
\leq  \v(t, s,z) \leq \sup_{\phi \in \Ac(t, x,q)} \mathbb{E}\left[\v^{*}(\theta^{\phi}, S^{t,s}_{\theta^{\phi}}, Z^{t,s,z,\phi}_{\theta^{\phi}})\right].
\end{align*}
\end{Proposition}

\proof Let $\Ac_{k}(t,x,z)$ be the set of controls $\phi$ satisfying   the additional  constraint $\#\{\tau^{\phi}_{i},i\ge 1\}\le k$ a.s., and let $\v_{k}$ be the corresponding value function, for $k\ge 1$. Then, it is not difficult to see that $\v_{k}$ is continuous, and  the arguments of  \cite{bouchard2011weak} imply that 
\begin{align*}
 \sup_{\phi \in \Ac_{k}(t, x,z)} \mathbb{E}\left[\v_{k}(\theta^{\phi}, S^{t,s}_{\theta^{\phi}},Z^{t,s,z,\phi}_{\theta^{\phi}})\right]
&\leq  \v_{k}(t,s, z)\\
& \leq \sup_{\phi \in \Ac_{k}(t, x,z)} \mathbb{E}\left[\v_{k}(\theta^{\phi}, S^{t,s}_{\theta^{\phi}}, Z^{t,s,z,\phi}_{\theta^{\phi}})\right]\\
&\le   \sup_{\phi \in \Ac(t, x,z)} \mathbb{E}\left[\v(\theta^{\phi}, S^{t,s}_{\theta^{\phi}}, Z^{t,s,z,\phi}_{\theta^{\phi}})\right].
\end{align*}
Since by definition $\v=\lim_{k\to \infty} \uparrow \v_{k}$ and $ \Ac(t, x,z)=\cup_{k\ge 1} \Ac_{k}(t, x,z)$, sending $k\to \infty$ in the above leads to the required result, recall Remark \ref{rem: v bounded HFT}.
\ep 
\vs2

  The partial differential equation associated to $\v$ is then at least formally given by 
 \begin{equation}\label{eq: edp HFT}
    \begin{aligned}
    					\min\left\{
					-\Lc \varphi -\Ic\varphi, \varphi -\Kc\varphi\right\} &= 0 \text{ on } [0, T) \times D_{S}\x D_{Z} \\
                        \min\left\{\varphi -U,\varphi -\Kc\varphi\right\}  &= 0 \text{ on } \{T\} \times D_{S}\x D_{Z},
     \end{aligned}
    \end{equation}
    in which $\Lc$ is the Dynkin operator associated to \eqref{eq: dyna S}:
    $$
    \Lc\varphi=\partial_{t}\varphi+\rho (\hat s- s) \partial_{s} \varphi +\frac12 \sigma^{2}\partial^{2}_{ss} \varphi.
    $$
 
To fully characterize the value function, we need the additional assumption, {similar to Assumption \ref{ass: comp MM}}. 

\begin{Assumption}\label{ass: comp  HFT} There exists a Borel  function  $\psi\in C^{1,2}([0,T]\x D_{S}\x D_{Z})$  such that 
\begin{enumerate}[\rm (i)]
\item $0\ge   \Lc\psi+ \Ic\psi$ on $[0,T)\x D_{S}\x D_{Z}$, 
\item $\psi-  \Kc\psi\ge \iota$ on   $[0,T]\x D_{S}\x D_{Z}$ for some $\iota>0$,
\item $\psi\ge   U$ on $\{T\}\x D_{S}\x D_{Z}$,
\item $\liminf\limits_{n\to \infty} (\psi/{\rm L})(t_{n},s_{n},z_{n})=\infty$ if  $|z_{n}|\to \infty$ as $n\to \infty$, for all $(t_{n},s_{n},z_{n})_{n\ge 1}\subset [0,T]\x D_{S}\x D_{Z}$.
\end{enumerate}
\end{Assumption}

\begin{Theorem}\label{thm : visco HFT} Let Assumption \ref{ass: thm visco MM} hold. Then, $\v_{*}$ (resp.~$\v^{*}$) is a viscosity supersolution (resp.~subsolution) of  \eqref{eq: edp HFT}. If moreover Assumption \ref{ass: comp  HFT} holds, then $\v$ is continuous on $[0,T)\x D_{Z}$ and is the unique viscosity solution of  \eqref{eq: edp HFT}, in the class $\L^{\rm exp}_{\infty}$.
\end{Theorem}

\proof  In view of  Proposition \ref{prop: DPP HFT}, the derivation of the viscosity super- and subsolution properties is very standard under Assumption \ref{ass: thm visco MM}, see e.g.~ \cite{bouchard2011weak}. 
As for uniqueness, this follows from a comparison principle that can be proved  in the class $\L^{\rm exp}_{\infty}$ by arguing as in the proof of Theorem \ref{thm : visco MM}. 
\ep 
 \vs2
 
 As for the MM problem, Assumption \ref{ass: comp  HFT} is easily checked when $J_{\circ}<\infty$. 

\begin{Remark}\label{rem : cond suff comp HFT} If $J_{\circ}<\infty$ and the supports of $\lambda(\cdot|p,q,c)$ and $\gamma(\cdot|p,q)$ are bounded, uniformly in $(p,q,c)\in ({\mathfrak d}\Z)^{2}\x\N^{2}\x \A_{\circ}$, then the function $\psi$ defined in Remark \ref{rem : cond suff comp HFT} also satisfies the requirements of   Assumption \ref{ass: comp HFT}, for $r$ large enough.
\end{Remark} 
\subsection{Dimension reduction, symmetries and numerical resolution}

As in Section \ref{subsec: numerics MM}, one can use specificities of the value function to reduce the complexity of the resolution of \eqref{eq: edp HFT}. First, the variable $(g,j)$ can be omitted, see Remark \ref{rem: dependence on g HFT}. 
Moreover, if Assumption \ref{ass: dep kernel by spread MM} holds, then one easily checks that 
$$
 {\bar \v}(t,s,p^{\mathfrak b},p^{\mathfrak b}+2\delta p,q^{\mathfrak b},q^{\mathfrak a},0,i,n^{\mathfrak b},n^{\mathfrak a},b^{\mathfrak b},b^{\mathfrak a},0)
$$
does not depend on $p^{\mathfrak b}$.  {The difference with the relation obtained in Section \ref{subsec: numerics MM} is du to \eqref{eq: def F} and  the fact that the HFT always hold a symmetric position in the stock and the futures (he is protected against evolutions of the mid-price).} The other symmetry relations described in Section \ref{subsec: numerics MM} do not hold because of the dependence on the process $S$. 
\vs2

Let us now turn to the definition of a numerical scheme for \eqref{eq: edp HFT}.  
Recall that, under Assumption \ref{ass : finite support}, the operators $\Ic$ and $\Kc$ are explicit. Hence, the only required discretization is in time and in the variable $s$. 
We shall consider separately the diffusion part and the obstacle part of the PDE. More precisely, we fix a time and a space grid  $\pi^{n}_{t}:=\{t^{n}_{i},i\le n_{t}\}$ and $\pi^{n}_{s}:=\{s^{n}_{i},i\le n_{s}\}$ where 
 $t^{n}_{i}=iT/{n_{t}}$ for $i\le n_{t}$ and $s^{n}_{i}=\underline s + i (\overline s-\underline s)/{n_{s}}$, $i\le {n_{s}}$.  Here, $\underline s$ and $\overline s$ are such that $D_{S}=[\underline s,\overline s]$, recall Remark \ref{rem : DS}, and $n:=(n_{t},n_{s})\in \N^{2}$. We next define the sequence of space domains $D_{Z}^k:=D_{Z}\cap [-k,k]^{11}$ for $k\ge 1$, and we let $\v^{k}_{n}$ be defined by  
  \begin{equation}\label{eq: edp MM num scheme 1}
    \begin{aligned}
    			&\v^{k}_{n} =\max\left\{\check \v^{k}_{n} ,  \Kc \check \v^{k}_{n}   \right\} = 0   \mbox{ on }\pi^{n}_{t}\x\pi^{n}_{s}\x   D_{Z}^k, \\
                        &\v^{k}_{n} -U  = 0 \text{ on } (\{T\}\times \pi^{n}_{s} \times D_{Z}^k)\cup (\pi^{n}_{t}\x \pi^{n}_{s}\x  (D_{Z}\setminus D_{Z}^k)).
     \end{aligned}
    \end{equation}
   Here, for $i\le n_{t}-1$ and $i'\le n_{s}$,
      \begin{equation}\label{eq: edp MM num scheme 2}
    \begin{aligned}
    &\check \v^{k}_{n}(t^{n}_{i},s^{n}_{i'},\cdot)=\E[\v^{k}_{n}(t^{n}_{i+1},{\rm p}_{n}(S^{t^{n}_{i},s^{n}_{i'}}_{t^{n}_{i+1}}),\cdot)]+\frac{T}{n}\Ic \v^{k}_{n}(t^{n}_{i+1},s^{n}_{i'},\cdot),   \mbox{ on }   D_{Z}^k,
     \end{aligned}
    \end{equation}
    where ${\rm p}_{n}$ is the left-hand side projection operator on $\pi^{n}_{s}$.

 Note that the above numerical scheme is not fully explicit as it requires to compute conditional expectations. This can however be easily performed either by a finite difference scheme or by Monte-Carlo techniques in a very classical way. Note in particular that the randomness in these conditional expectations only comes from a one dimensional factor.   
   
\begin{Proposition}\label{prop: conv vnk HFT} Let Assumptions \ref{ass: comp HFT} and \ref{ass : finite support} hold, then  the sequence   $(\v^{k}_{n})_{n\ge 1}$ converges pointwise to $\v$ as $n_{t},n_{s},k\to \infty$.
\end{Proposition}

\proof Note that Assumption \ref{ass: comp HFT} ensures that comparison holds for    \eqref{eq: edp HFT} in the class $\L^{\rm exp}_{\infty}$, see e.g., \cite[Proposition 5.1]{baradel2016optimal}. Thus, as for Proposition \ref{prop: conv vnk MM}, the result is an easy consequence of {the stability result of} \cite{barles1991convergence}. \ep 

\subsection{Approximate optimal controls}

The optimal control can be numerically estimated as in Section \ref{subsec: numerics MM}. 
We first extend $(\v^{k}_{n},\check \v^{k}_{n})$ to $\pi^{n}_{t}\x D_{S}\x D_{Z}$ by setting $(\v^{k}_{n},\check \v^{k}_{n})(\cdot,s,\cdot):=(\v^{k}_{n},\check \v^{k}_{n})(\cdot,{\rm p}_{n}(s),\cdot)$. 
Then, we choose a measurable map $\hat c_{n}^{k}(t^{n}_{i},\cdot)$  such that 
\begin{align*}
&\hat c_{n}^{k}(t^{n}_{i},\cdot)\in {\rm arg}\max \{\Kc^{c} \check \v^{k}_{n}(t^{h}_{i},s,\cdot),\;c\in \A(\cdot)\}\;,\; \mbox{ on }   D_{S}\x D_{Z}^k\\
&\hat c_{n}^{k}(t^{n}_{i},\cdot)=0 \mbox{ on } D_{S}\x(D_{Z}\setminus D_{Z}^k),
\end{align*}
and define the sequence of stopping times 
$$
\hat \tau^{n,k}_{j+1}:=\min\{t^{n}_{i}: i\ge 0, t^{n}_{i}>\hat \tau^{n,k}_{j},(\v^{k}_{n}-\Kc^{\hat c_{n}^{k}}\check \v^{k}_{n}(t^{n}_{i},\cdot)) (t^{n}_{i}, S_{t^{n}_{i}},\hat Z_{t^{n}_{i}-})=0\},\;j\ge 0,
$$
with $\hat \tau^{n,k}_{0}:={0-}$, and in which $\hat Z^{n,k}=(\hat P^{n,k},\hat Q^{n,k},\hat X^{n,k})$ is defined as in \eqref{eq: dyna P Q X}-\eqref{eq: dyna S} for the initial condition $Z_{0-}$ and the control associated to $\hat \phi^{k}_{n}:=(\hat \tau^{n,k}_{i},\hat c_{n}^{k}(\hat \tau^{n,k}_{i},S_{\hat \tau^{n,k}_{i}},\hat Z^{n,k}_{\hat \tau^{n,k}_{i}-}))_{i\ge 1}$ in a Markovian way.  
\vs2
Again, this provides a sequence of controls that is asymptotically optimal. 

\begin{Proposition} Let the conditions of Proposition \ref{prop: conv vnk HFT} hold. Then,
$$
\lim_{k\to \infty}\lim_{n_{t},n_{s}\to \infty} \E[U(S_{T},\hat Z^{n,k}_{T})]=\v(0,S_{0},Z_{0-}), 
$$
in which the limit is taken along sequences $n$ such that $n_{t}^{2}n_{s}^{-1}\to 0$.
\end{Proposition}

\proof 
Let $\gamma(p,q):=\int d\beta(c|p,q)$ and recall that $\gamma$ is uniformly bounded by assumption. The family $\{(\v^{k}_{n},\check \v^{k}_{n})/{\rm L}\}_{k,n\ge 1}$ is uniformly bounded, compare with Remark \ref{rem: v bounded HFT}. 
Also note that $(s,z)\in {D_{S}}\x D_{Z}^{k}\mapsto U$ is $C_{k}$-Lipschitz, for some $C_{k}>0$ that only depends on $k$. By induction {(recall that the component $s$ is projected on $\pi^{n}_{s}$)}, 
$$
|\check \v^{k}_{n}(\cdot,s,\cdot)-\check \v^{k}_{n}(\cdot,s',\cdot)|\le C'_{k}[|s-s'|+n_{t}O(n_{s}^{-1})], \mbox{ for all } s,s'\in D_{S},\mbox{ on } D_{Z}^{k},
$$
 for some $C'_{k}>0$ that does not depend on  $n{\in \N^{2}}$. Since $S$ is   $1/2$-H\"older in time in ${\mathbf L}_{2}$, together with Assumption \ref{ass : finite support}, this implies that  
 \begin{align*}
\Kc^{c}\Ic \check \v^{k}_{n}(t^{n}_{i+1},s^{n}_{i'},z)&=\Kc^{c}\E[\Ic\check \v^{k}_{n}(t^{n}_{i+1},S^{t^{n}_{i},s^{n}_{i'}}_{t^{n}_{i+1}},z)]+ O_{k}(n_{t}^{-\frac12})+n_{t}O_{k}(n_{s}^{-1}),\\
&= \E[\check \v^{k}_{n}(t^{n}_{i+1}, S^{t^{n}_{i},s^{n}_{i'}}_{t^{n}_{i+1}},Z^{t^{n}_{i},z,c}_{t^{n}_{i+1}})]+O_{k}(n_{t}^{-\frac12})+n_{t}O_{k}(n_{s}^{-1})\;\;\mbox{for $z\in D^{k}_{Z}$, $c\in \A(z)$,}
\end{align*}
in which the exponent $c$ in $Z^{t^{n}_{i},z,c}$ means that the impulse $c$ is given at $t^{n}_{i}$, and we use Remark \ref{rem : number of jumps in small time} again. In the above, $O_{k}(\xi)$ is a function, that may depend on $k$, but such that $\xi \in \R\setminus\{0\}\mapsto|O_{k}(\xi)|/\xi$ is bounded in a neighborhood of $0$.  
By the arguments already used in the proof of Proposition \ref{prop : conv control opti MM}, it follows that 
\begin{align*}
  \v^{k}_{n}(t^{n}_{i},s^{n}_{i'},z)=& \E[\v^{k}_{n}(t^{n}_{i+1},S^{t^{n}_{i},s^{n}_{i'}}_{t^{n}_{i+1}},Z^{t^{n}_{i},z,0}_{t^{n}_{i+1}})]\vee \max_{c\in \A(z)} \E[\v^{k}_{n}(t^{n}_{i+1},S^{t^{n}_{i},s^{n}_{i'}}_{t^{n}_{i+1}},Z^{t^{n}_{i},z,c}_{t^{n}_{i+1}})]\\
  &+ o_{k}(n_{t}^{-1})+n_{t} O_{k}(n_{s}^{-1}), 
\end{align*}
and therefore 
\begin{align*}
\v^{k}_{n}(0,S_{0},Z_{0-})&=\E[U(S_{\theta^{n}_{k}},\hat Z^{n,k}_{\theta^{n}_{k}})]+ o^{k}_{n}(1)+O_{k}(n_{t}^{2}n_{s}^{-1})\\
&=\E[U(S_{T},\hat Z^{n,k}_{T})]+ o^{k}_{n}(1)+O_{k}(n_{t}^{2}n_{s}^{-1})+o_{k}(1)
\end{align*}
by induction, in which $o^{k}_{n}(1)$ goes to $0$ as $n\to \infty$, $o_{k}(1)$  goes to $0$ as $k\to \infty$,  and $\theta^{n}_{k}$ is as in the proof of Proposition \ref{prop : conv control opti MM}. It remains to appeal to Proposition \ref{prop: conv vnk HFT}. \ep 

\subsection{Numerical experiments}\label{sec: numeric HFT}

We use the same model as the one described in Section \ref{subsec: num expe MM}. As for the new parameters, we take      {$\hat s=0$, $\rho=50$ and $\sigma=0.2$}, in particular the mean reversion parameter is taken to be large. 
 \begin{figure}[h!]
 \includegraphics[width=\textwidth,height=0.86\textheight]{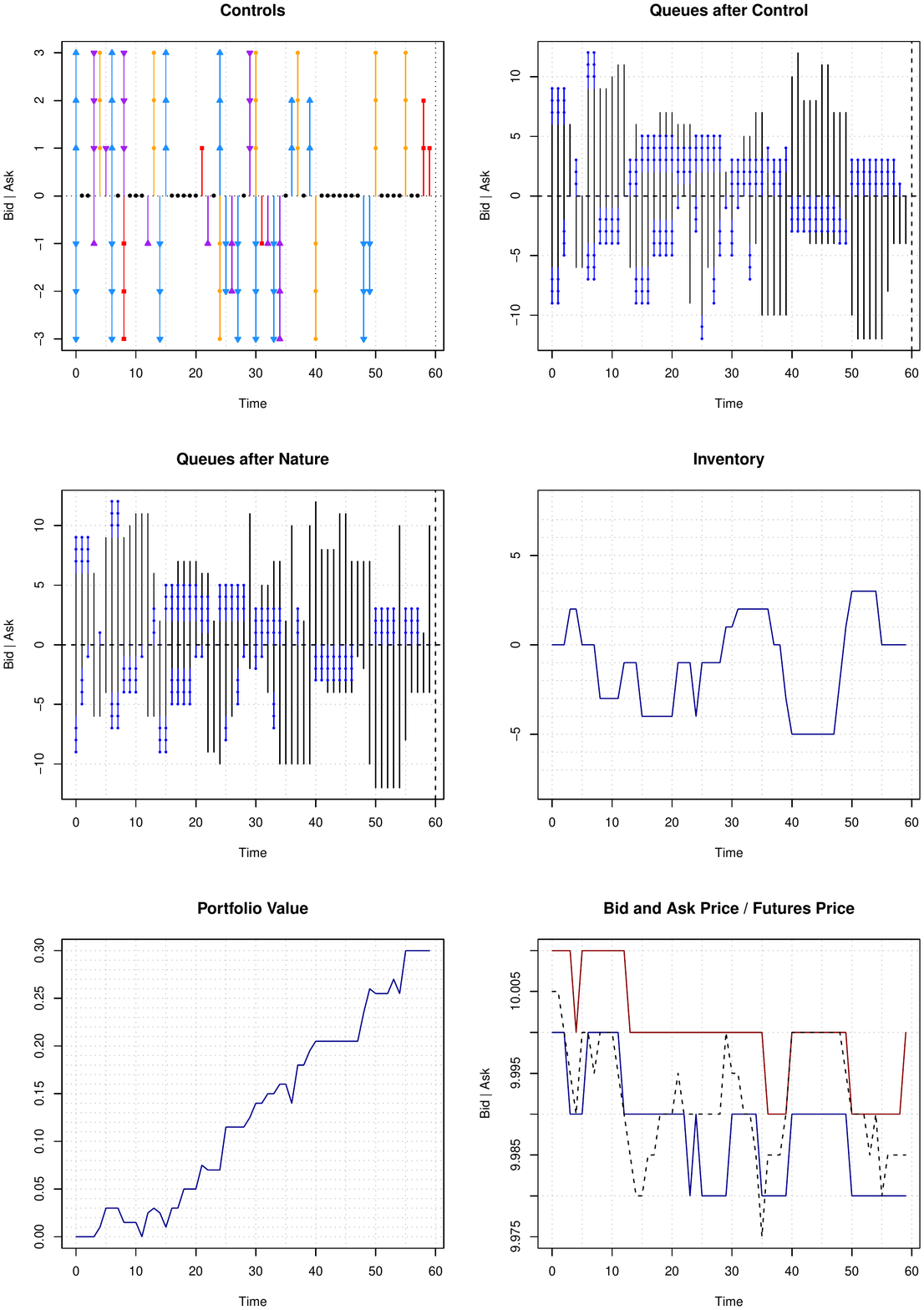}
 \caption{Simulated path of the optimal strategy of the High Frequency Trader.} 
\label{fig:HFT}
\end{figure}
\newpage
We approximate the behavior of the spread process $S$ by a trinomial tree based on the transition probabilities associated to the diffusion \eqref{eq: dyna S}, so that the expectation in \eqref{eq: edp MM num scheme 2} can be computed explicitly. More precisely, we consider a centered 6 points grid with mesh equal to half a tick (i.e.~$0.005$). 

The graphics in Figure \ref{fig:HFT} have the same interpretation as in   Section \ref{subsec: num expe MM} except that we now also provide the evolution of the Futures process $F$, this is the dashed line in the bottom right graphic. 
\\
Remember that the HFT does not gain from the evolution of the mid-price, as her position in stocks in always covered by a symmetric position in the futures. She  only gains from the evolution of the spread process $S$ or from the bid-ask spread of the stock if   $S$ does not move.  Not surprisingly, her behavior is quite different from the one of the MM described in Section \ref{subsec: numerics MM}. \\
 As the MM would do, she first positions herself on the limits in a symmetric ways, because the spread with the future is $0$. She then tries selling at time $4$ by sending a limit sell order in the spread to buy the futures whose price (compared to the mid-price of the stock)  is gone low. She is starting to play on the stock-futures spread. She follows this strategy until time $14$.  She plays in a more symmetric way after this until time $35$, with a slight tendency to resume her inventory. At time $35$,  she decides to clearly sell stocks again and buy the futures whose price is again very low. From time $40$ on, she inverts her position on the book to try buying the stock and thus sells the futures whose price has gown up. She finally inverts her position again at time $50$ when the futures price goes back to the mid price, to liquidate her position on the pair.  
 \\
 Compared to the positions of the MM in Figure \ref{fig:MM75}, her positions in one direction are much more clear, and are clearly driven by the stock-futures spread. The HFT also does not seem to try to control the stock's mid-price as the MM did, again this is because it does not matter for her. 

\begin{figure}[h!]
\centering
 \includegraphics[width=0.5\textwidth]{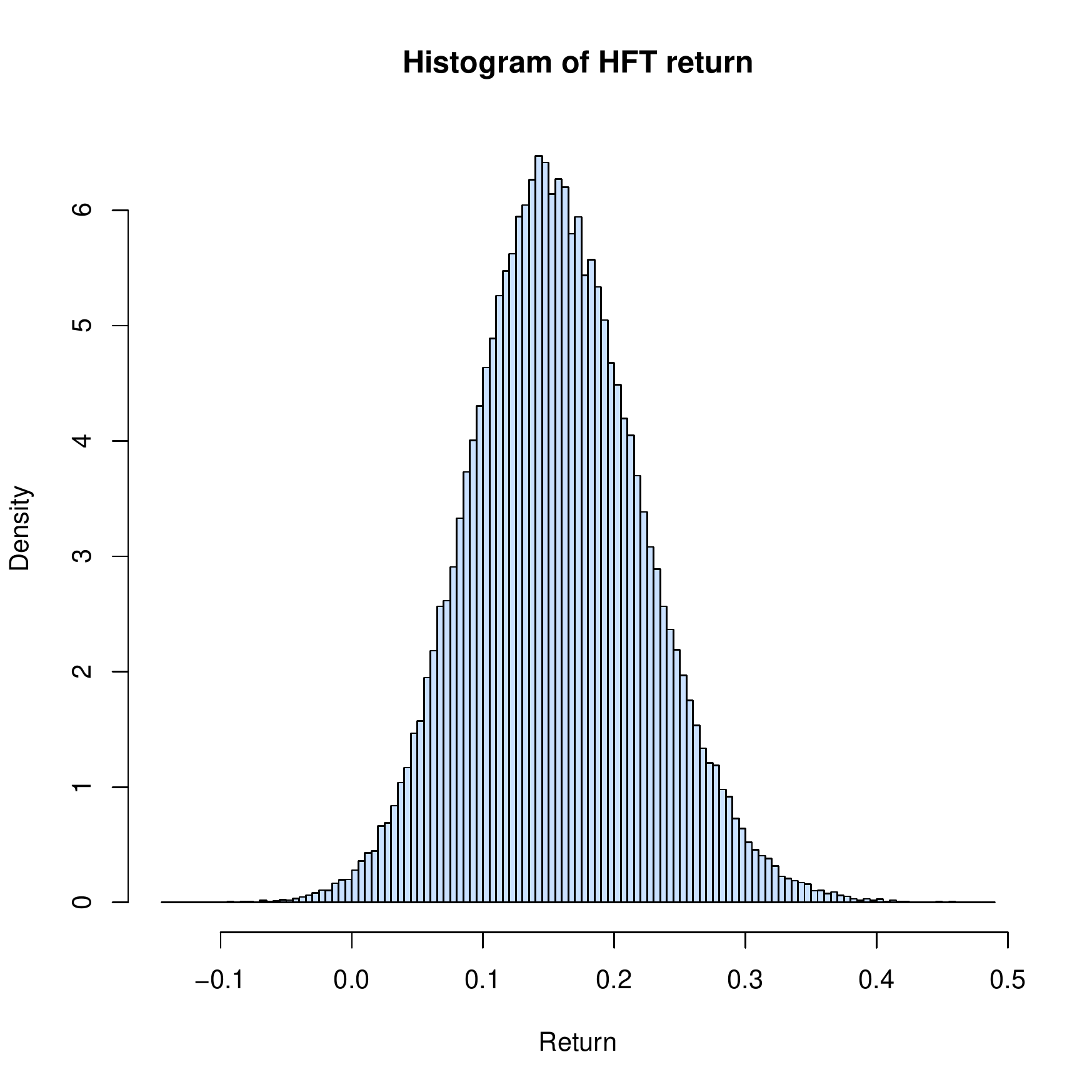}
 \caption{Density estimation of the gain made by  the High Frequency Trader.} 
\label{fig:HistHFT}
\end{figure}

In Figure \ref{fig:HistHFT}, we provide an estimation of the density of the gain of the HFT based on $10^{5}$ simulated paths.    

 \section{Institutional broker strategies for portfolio liquidation}\label{sec:InstitutionalBroker}
\def\vc{\vartheta}
\def\w{{\rm w}}

We now turn to the Institutional Broker problem. We consider in the following the two mostly used strategies for buying/selling a block of stocks. We focus on the buying side, selling being performed in a symmetrical way.

\subsection{Volume strategy}\label{subsec: volume strategy}

We first consider a simple volume strategy.  The aim of the broker is to buy ${\rm I_{0}}$ stocks. By convention, we set $I_{0}=-{\rm I}_{\circ}$. She fixes a participation rate $f\in (0,1)$. It corresponds to the percentage of the total volume of trades that should correspond to trades done by the broker between the initial time $0$ and the time at which the ${\rm I_{0}}$ stocks are bought. Therefore, she should buy $f/(1-f)$ of the total of the trades of the other market participant. To do so,   she considers subintervals  $[t_{i},t_{i+1}]$ of $\R_{+}$. At the beginning of each of this subintervals she estimates  the conditional {probability} ${\rm p}_{i,i+1}$ that a trade arrives at the bid, given that a trade arrives and given the {order book} at $t_{i}$. Then, she puts a limit bid order of size $(f/(1-f))Q^{\bk}_{t_{i}}/{\rm p}_{i,i+1}$ at $t_{i}$. When {all these} units are executed, she adds new limit orders to keep her position to a level at least equal to   $(f/(1-f))Q^{\bk}_{t_{i}}/{\rm p}_{i,i+1}$, or do nothing if this level is still achieved, {and so on}. In the meantime, she compares her realized volume $\Delta I:=I-I_{0}$ to the volume  $\tilde v(0,\cdot)+\Delta I$ realized on the market from time $0$ on. Given a threshold $\delta_{I}>0$, she stops adding limit orders and cancels her orders already at the bid if $\Delta I(1-f)>f \tilde v(0,\cdot)+\delta_{I}$ and waits until $\Delta I(1-f)\le f \tilde v(0,\cdot)+\delta_{I}$. If $\Delta I(1-f)<f \tilde v(0,\cdot)-\delta_{I}$, then she sends an aggressive order so as to turn to a position $\Delta I(1-f)\ge f \tilde v(0,\cdot)-\delta_{I}$ as soon as possible.  She stops trading when $I\ge 0$.
\vs2

\begin{figure}[h!]
\centering
 \includegraphics[width=0.8\textwidth,,height=0.5\textheight]{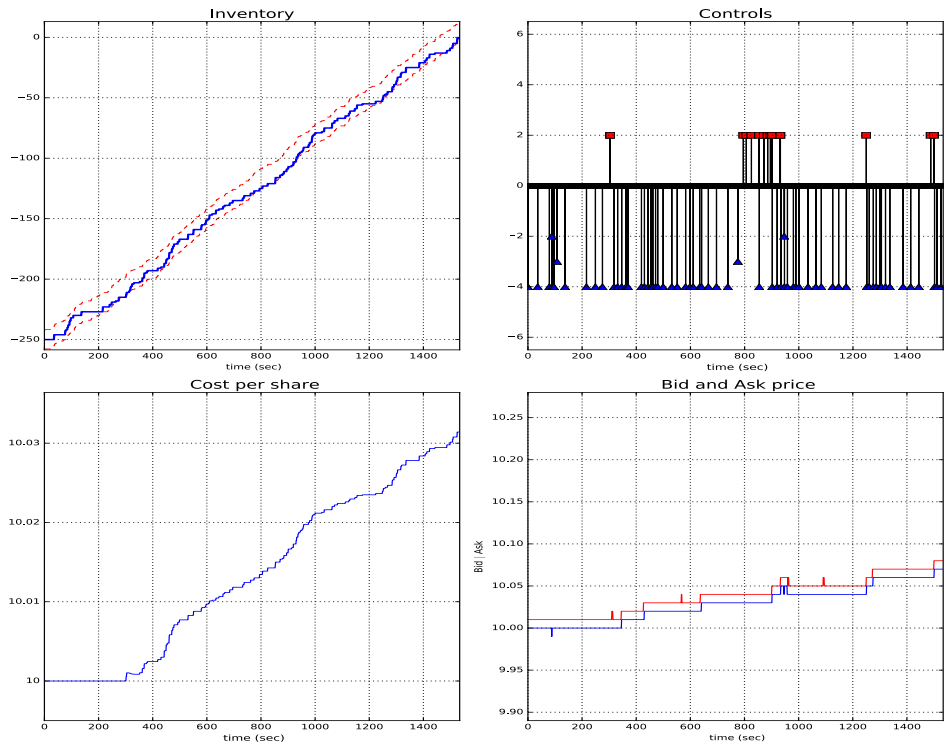}
 \caption{Simulated path of the volume strategy.} 
\label{fig:volume}
\end{figure}

For our numerical experiment, we take the same model as in Section \ref{subsec: num expe MM}, we consider a participation rate $f=0.2$ and simply take ${{\rm p}_{i,i+1}=1/2}$. The time intervals $[t_{i},t_{i+1}]$ have a length of $60$ seconds and the time step is $1/2$ second.  We take $\delta_{I}=4$ (i.e.~$2$ ATS). The number of units to buy is  {${\rm I}_{\circ}=250$ units.} A simulated path of the strategy is provided in Figure \ref{fig:volume}. In the top left graphic, the dashed lines correspond to the target volume $\pm \delta_{I}$, while the solid curve is the volume traded by the IB. The top right graphic gives the control of the IB: lines with inward pointing arrows are limit orders, lines with squares are market orders. The bottom left graphic provides the evolution of the average price at which stocks have been bought by the IB from time $0$ on. The bottom right graphic gives the evolution of the bid and ask prices.   One can see that this very simple strategy is quite efficient in the sense that only a limited number of market orders are sent. On the other hand, the systematic position of the IB at the bid limit creates an important imbalance that contributes to push up the price.

\begin{figure}[h!]
   \centering
 \includegraphics[height=0.3\textheight]{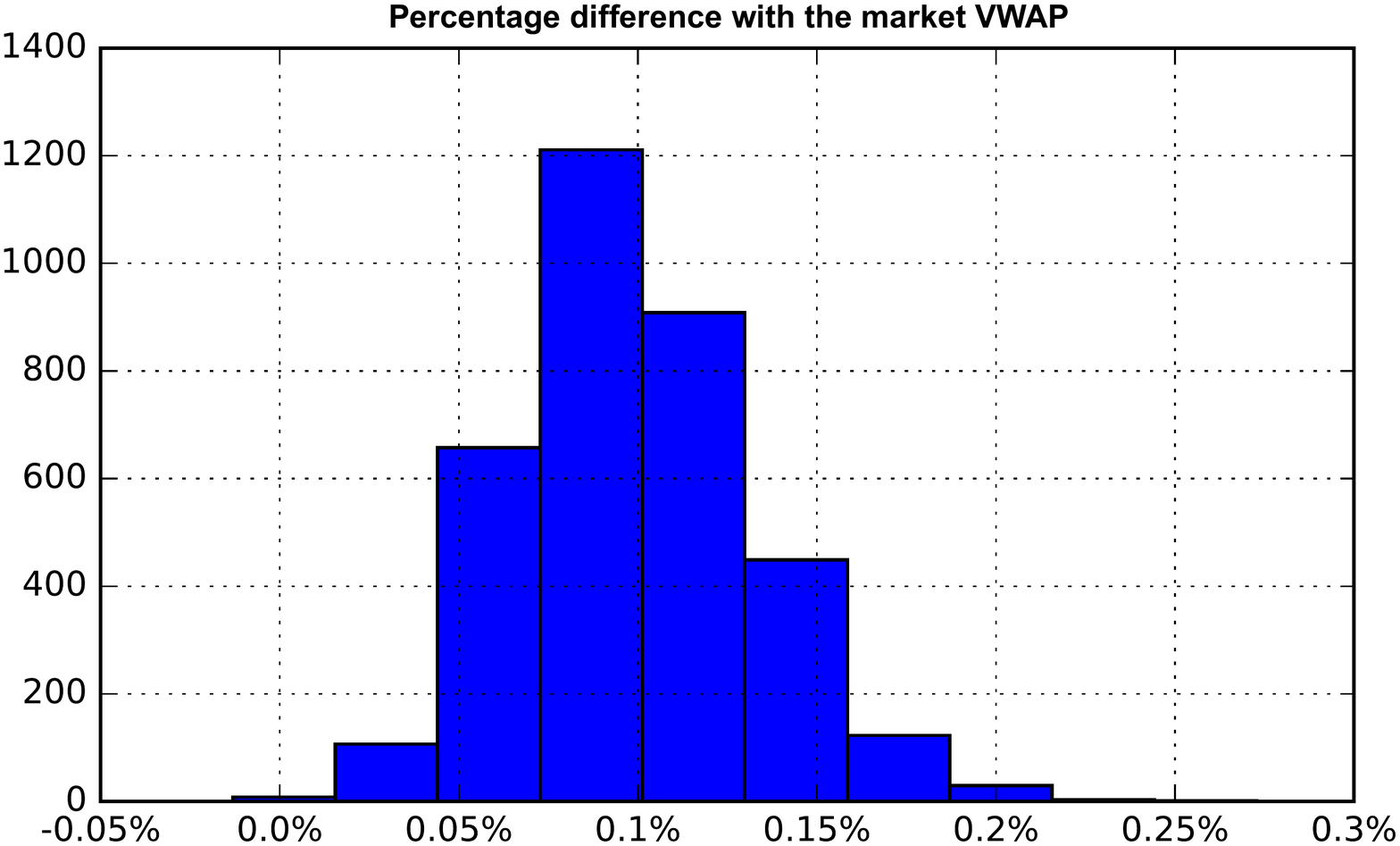}
 \caption{{Histogram} of the VWAP percentage error with respect to the VWAP of the market, for the Volume strategy.} 
\label{fig:HistVolume}
\end{figure}

 Figure \ref{fig:HistVolume} provides an histogram of the relative error (in $\%$) of the VWAP obtained by this strategy with respect to the VWAP realized at the level of  the whole market\footnote{Namely, $({\rm VWAP}_{\rm Vol}-{\rm VWAP}_{\rm Market})/{\rm VWAP}_{\rm Market}$ in which ${\rm VWAP}_{\rm Vol}$ is the VWAP obtained by the IB by playing his volume strategy.}. It is based on $10^{4}$ simulated paths. One can see that this average price is typically slightly higher than the VWAP of the market.

\newpage


\subsection{VWAP}

We now present a VWAP (Volume Weighted Average Price) based trading strategy frequently practiced by institutional brokers.  Suppose that, at initial time $0$, an institutional broker decides to buy a quantity ${\rm I}_0\in \N$ of a tradable asset using a VWAP based strategy, i.e.~by trying to obtain  an average execution cost not more than 
 \begin{equation}
\text{VWAP}_T=\frac{\w_{T}}{v(0,T)}\;\mbox{ where } \w:=\int_{0}^\cdot v_t P_t dt\mbox{ and }v(\cdot,T):=\int_\cdot^T v_sds,
\label{eq:vwap}
\end{equation}
in which $v$ is a deterministic non-negative continuous process such that $\int_0^T v_tdt>0$, which represents the trading volume of the market, and $P$ models the stock price.\footnote{We take it deterministic for simplicity. In practice, it generally corresponds to a market volume curve estimated by the broker.} 

\subsubsection{Abstract continuous time resolution}\label{sec: VWAP continuous time}

Following the seminal work of Almgren and Chriss \cite{almgren1999value}, see also \cite[Section 4.4]{gueant2016financial},  we first consider the idealized world in which trading is done continuously at a bounded intensity, the  trading speed $\vc$, taken as  a process in the class $\Ac$ of non-negative processes adapted  to the (completed) filtration generated by $W$.   She assumes that the  dynamics of the asset reference price $P$ has a permanent linear price impact generated by the agent's trading activity, see \cite{almgren1999value}. More precisely, she assumes that the stock price evolves according to 
\begin{equation}\label{eq: dyna P vwap}
dP_t = \beta \vartheta_tdt+\sigma dW_t,
\end{equation}
with $\beta,\sigma>0$, 
while her inventory $I$ follows the dynamics
\[
dI_t = \vartheta_t dt
\] 
with the initial condition $I_{0}=-{\rm I}_{0}$, meaning that she is short at $0$ of  the  ${\rm I}_{0}$ stocks she has to buy on $[0,T]$.
 
She also assumes that her  wealth $G$ is affected by a temporary linear market impact $\kappa \vartheta$, for some $\kappa>0$, i.e.
\begin{equation}\label{eq: dyna G vwap}
dG_{t}=-\vc_t \big[P_t + \kappa \vc_t\big]dt.
\end{equation}
 
Her goal is to maximize over $\vartheta \in \Ac$ the expected utility 
\begin{align*}
 \Esp{-\exp[-\eta\{G_T + I_T(P_T-\tilde{\kappa}I_T)-{\rm I}_0\frac{\w_{T}}{v(0,T)}\}]}   \numberthis \label{Eq:ValueFunctionAgent} ,
\end{align*}
for some $\tilde \kappa>0$ which represents a penalty in case the  inventory does not match $0$ at  $T$. 

\vs2

 This type of  problems has been widely studied, see the book  \cite{gueant2016financial} for references. In the present form, it can be solved explicitly by using a verification argument based on the explicit solution of the  Hamilton-Jacobi-Bellman equation corresponding to the value function $(t,p,i,g,w)\in [0,T]\x \R^{4}\mapsto \v(t,p,i,g,w)$ associated to the initial condition $(P_{t},I_{t},G_{t},\w_{t})=(t,p,i,g,w)$: 
 $$
 0=\sup_{u\ge 0} \left(\partial_{t}\vp +\frac12 \sigma^{2}\partial^{2}_{pp} \vp+pv_{t}\partial_{w}\vp +u \left(\beta \partial_{p}\vp+\partial_{i}\vp-p\partial_{g}\vp\right)-u^{2}\kappa \partial_{g}\vp\right)
 \;\mbox{ on } [0,T)\x \R^{4}
 $$
 with terminal condition 
 $$
 \vp(T,p,i,g,w)=-e^{-\eta\{g + i(p-\tilde{\kappa}i)-\bar m_{0}w\}} \;\mbox{ for } (p,i,g,w)\in \R^{4},
 $$ 
 where 
 $$
 \bar m:={\rm I}_{0}/v(0,T).
 $$
 To simplify the above, we first write $\v$ is the form
 $$
  \v(t,p,i,g,w)=e^{-\eta(g-{\bar m}w- pv(t,T){\bar m})}\bar \v(t,p,i)\mbox{ with } \bar \v(t,p,i):=\v(t,p,i,0,0), 
 $$
so that $\bar \v$ formally solves 
 $$
0=\partial_{t}\vp +\frac12 \sigma^{2}(\partial^{2}_{pp} \vp+2\eta v(t,T)\bar m\partial_{p}\vp+\eta^{2}v(t,T)^{2}\bar m^{2}\vp)-\frac{(\beta \partial_{p}\vp+\beta \eta v(t,T)\bar m\vp+\partial_{i}\vp+\eta p\vp)^{2}}{4\kappa \eta\vp}
 $$
 on $[0,T)\x \R^{4}$, 
 with terminal condition 
 $$
 \vp(T,p,i,g,w)=-e^{-\eta\{ i(p-\tilde{\kappa}i)\}} \;\mbox{ for } (p,i,g,w)\in \R^{4}.
 $$
One possible  solution is given by 
 $$
 \bar \v(t,p,i)=-e^{-\eta\{ i(p-\tilde{\kappa}i)\}}e^{h_{0}(t)+h_{1}(t)i+h_{2}(t)i^{2}}
 $$
in which $h_{0}, h_{1}$ and $h_{2}$ solve\footnote{{Just insert the above in the PDE of $\bar \v$ and match the orders in the $i$ variable.}} on $[0,T)$
\begin{align*}
\partial_{t}h_{0}&=-\frac12 \sigma^{2} (\eta v(t,T)\bar m)^{2}+\frac{(h_{1}+\beta \eta v(t,T) \bar m)^{2} }{4\kappa \eta}\\
\partial_{t}h_{1}&=\sigma^{2} \eta^{2} v(t,T)\bar m+(h_{1}+\beta\eta v(t,T) \bar m)\frac{-\beta \eta+2\eta \tilde \kappa+2h_{2}}{2\kappa \eta}\\
\partial_{t}h_{2}&=-\frac{\eta^{2}\sigma^{2}}{2}+\frac{\left[-\beta \eta+2\eta \tilde \kappa+2h_{2}\right]^{2}}{4\kappa \eta}
\end{align*} 
with $h_{0}(T)=h_{1}(T)=h_{2}(T)=0$. The solution of the last equation is of the form 
$$
h_{2}(t)=\frac{1}{c_{1}+c_{2}e^{c_{3}(T-t)}}-c_{4}
$$
in which the constants $c_{1},\ldots, c_{4}$ can be computed explicitly by using the above differential equation and the terminal condition $h_{2}(T)=0$. Namely, set 
$$
a_{0}:=-\frac{\eta^{2}\sigma^{2}}{2}+\frac{\eta(2\tilde \kappa-\beta)^{2}}{4\kappa }, \; a_{1}:=\frac{2\tilde \kappa-\beta}{\kappa },\;a_{2}=\frac{1}{\kappa \eta}
$$
and let $y_{\circ}$ be a root\footnote{The function $h_{2}$ does not depend on the choice of the root.} of 
$$
(4 a_{0}a_{2}-a_{1}^{2})y^{2}+(a_{1}^{2}-4a_{0}a_{2})y+a_{0}a_{2}=0,
$$ 
then 
$$
c_{3}=\frac{a_{1}}{1-2y_{\circ}},\;c_{4}=\frac{a_{0}}{(1-y_{\circ})c_{3}},\;c_{1}=\frac{y_{\circ}}{c_{4}},\;c_{2}=\frac1{c_{4}}-c_{1}.
$$
Existence of $y_{\circ}$ is guaranteed since 
\begin{align*}
a_{1}^{2}-4a_{0}a_{2}=\frac{\eta^{2}\sigma^{2}}{2}> 0.
\end{align*}
 Then,   $h_{0}$ and $h_{1}$ are fully characterized.

An easy verification argument shows that this is actually the correct function and that the  optimal control policy is given by 
$$
\hat \vc=\hat v(\cdot,I)
$$
where 
\begin{align*}
\hat v{(t,i)}&:=\frac{\beta\partial_{p}\v(t,p,i,g,w) +\partial_{i}\v(t,p,i,g,w)-p\partial_{g} \v(t,p,i,g,w)}{2\kappa \partial_{g}\v(t,p,i,g,w)}\\
&=\frac{\beta[ i- v(t,T){\bar m}]  -2 \tilde \kappa i-[h_{1}(t)+2h_2(t)i ]/\eta}{2\kappa }.
\end{align*}

\subsubsection{Strategy in practice and simulations}

 In practice, this optimal strategy can not be implemented within an order book. We therefore consider a ``discrete'' version. In this version, we assume that the IB tries to keep her inventory $I$ equal to the optimal inventory $\hat I:=\int_{0}^{\cdot} \hat \vc_{t}dt-{\rm I}_{\circ}$. To do this, she fixes a time grid $\{t_{i},i\le n\}$ with $t_{0}=0$ and $t_{n}=T$. At time $t_{i}$, she estimates that she has to execute a volume of $V_{i,i+1}:=\int_{t_{i}}^{t_{i+1}}\hat v({t, I_{t_{i}}})dt$ on $[t_{i},t_{i+1}]$ while the volume of the market will be $\tilde V_{i,i+1}=\int_{t_{i}}^{t_{i+1}} v_{t}dt$. Then, she follows a volume strategy with threshold $\delta_{I}=+\infty$, see Section \ref{subsec: volume strategy} above, on $[t_{i},t_{i+1}]$ with a target participation rate of  $f= V_{i,i+1}/(\tilde V_{i,i+1}+V_{i,i+1}) $. When $\Delta I:=I-I_{0}>\hat I + \bar \delta_{I}$,  for some $\bar \delta_{I}>0$, orders are canceled and she waits until 
  $\Delta I\le \hat I + \bar \delta_{I}$.   In the case where   $ \Delta I< \hat I-\bar \delta_{I}$, she puts market orders to reduce to the situation  $ \Delta I\ge  \hat I-\bar \delta_{I}$ as soon as possible.

 For our numerical experiment, we take the same configuration as in Section \ref{subsec: volume strategy}, with $ \bar \delta_{I}={4}$, i.e.~ ${2}$ ATS.   The optimization of the VWAP strategy is done with a time horizon of $30$ minutes and a flat volume curve (so that the control does in fact not depend on it). The number of units to buy is  {${\rm I}_{\circ}=250$ units.} The additional parameters\footnote{The coefficients $\beta$ and $\kappa$ are estimated for our book dynamic. Given priors with simulate a bunch of paths and estimated them by a moment matching approach based on \eqref{eq: dyna P vwap}-\eqref{eq: dyna G vwap}. We then use the updated values to simulate a new bunch of paths and we re-estimate them. And so on, until convergence. } are set to  {$\eta=1$, $\sigma=0.2$},  {$\beta=0,0004$,  $\kappa=0.003$ and $\tilde \kappa=\kappa*60$. The latter corresponds to the cost incurred when buying the remaining shares $I_{T}$ in $1$ additional minute after $T$, at a flat intensity in the theoretical continuous time model of Section \ref{sec: VWAP continuous time}.  The volume intensity $v$ corresponds to $0.6$ ATS per second, i.e.~$1.2$ units per second. }
 \\
 
\begin{figure}[h!]
   \centering
 \includegraphics[width=0.8\textwidth,height=0.5\textheight]{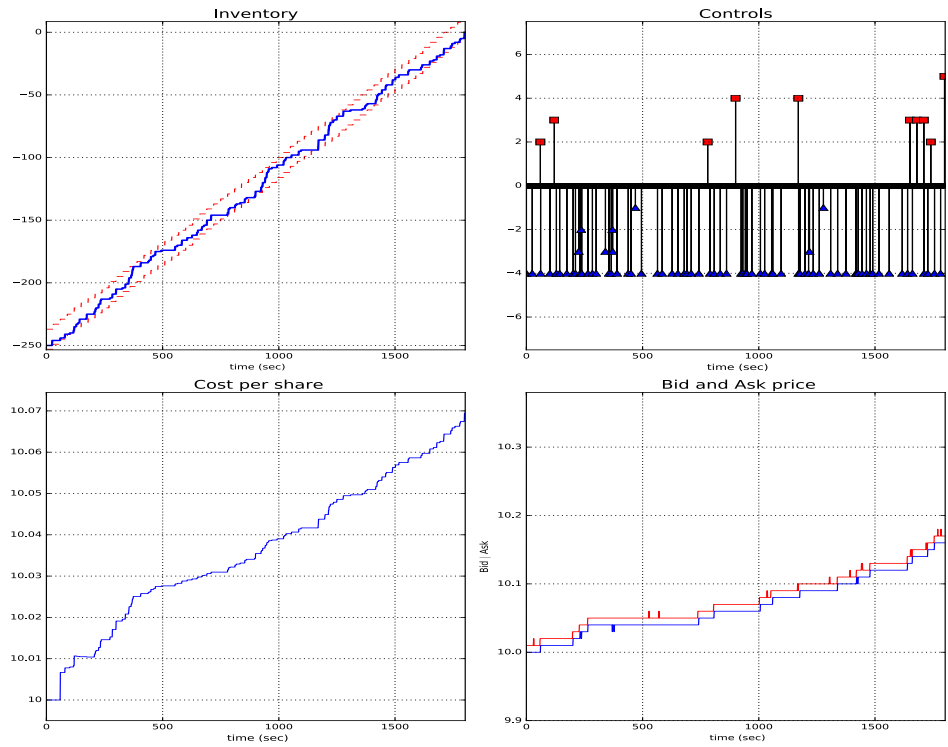}
 \caption{Simulated path of the VWAP strategy.} 
\label{fig:vwap}
\end{figure}

 The interpretation of the different graphics in Figure \ref{fig:vwap} is the same as  in Section \ref{subsec: volume strategy}, except that the dashed lines in the top left graphic correspond now to the optimal VWAP trading curve $\pm 2$ ATS.   Again, we see that only a limited number of market orders  were needed  to be sent, but that the imbalance created by the  {VWAP trading algorithm} drives the price up.  
  
\begin{figure}[h!]
   \centering
 \includegraphics[height=0.3\textheight]{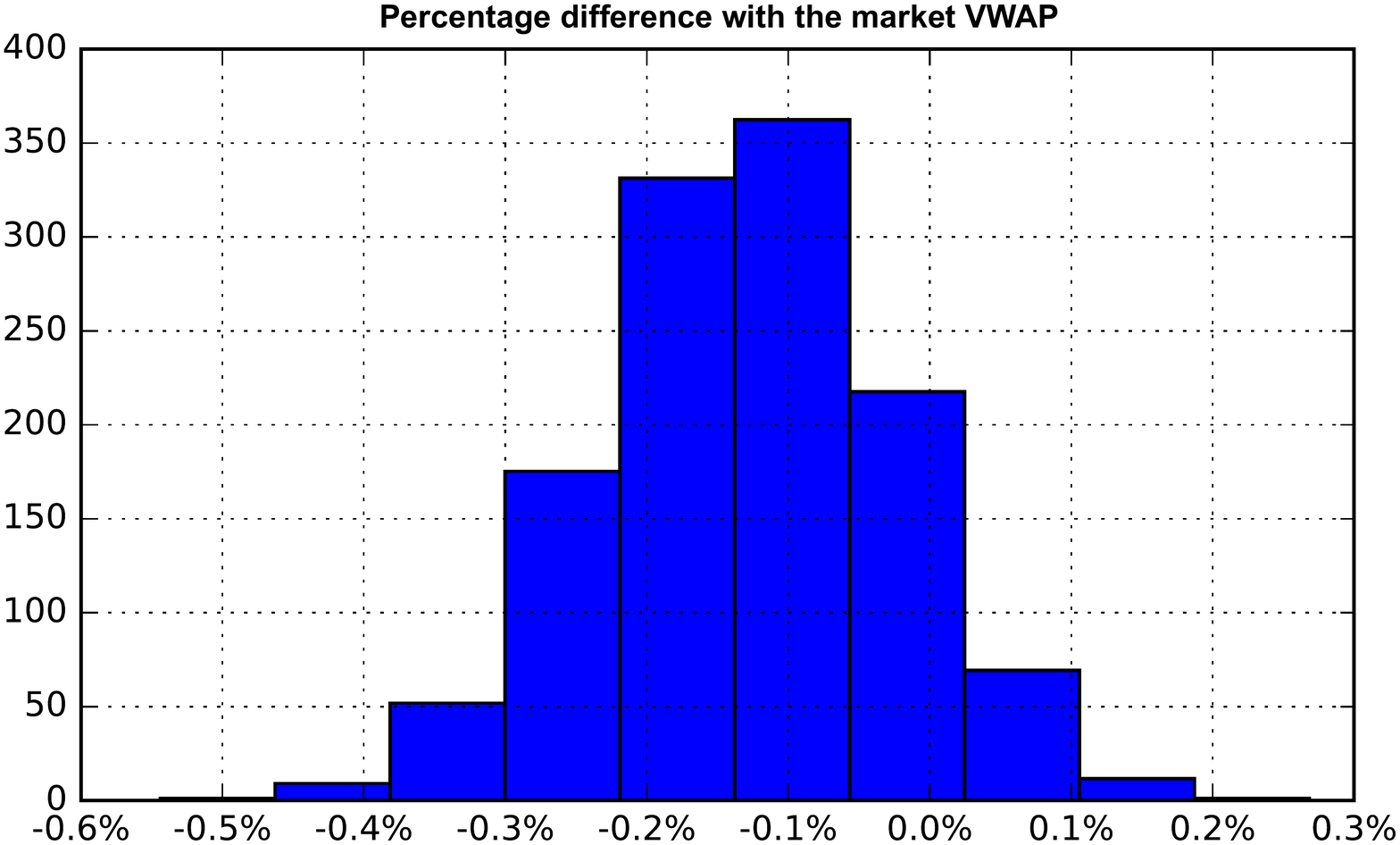}
 \caption{{Histogram} of the percentage error with respect to the VWAP of the market, for the VWAP strategy.} 
\label{fig:Histvwap}
\end{figure}

  Figure \ref{fig:Histvwap} provides an histogram of the relative error (in $\%$) of the VWAP obtained by this strategy with respect to the VWAP of the whole market\footnote{Namely, $({\rm VWAP}_{\rm MM}-{\rm VWAP}_{\rm Market})/{\rm VWAP}_{\rm Market}$, in which ${\rm VWAP}_{\rm MM}$ is the VWAP obtained by the IB by playing his optimal VWAP strategy}.  It is based on $10^{4}$ simulated paths. One can see that he actually typically  performs better than the market. Not surprisingly, this strategy performs better than the volume strategy in terms of VWAP.

\newpage


\section{Simulation of the full market }\label{sec : Simulation of the global market}
 
We now provide an illustration of possible interactions of several market participants. Again, a more realistic and general  study is left for future research. 

\vs2

The prior and parameters  of the participants are the same as the one that have been used in the above numerical experiments. More precisely, we consider one MM, {one} HFT and four IB. Among the IB, two are playing volume strategies (a buy and a sell) and two are playing WVAP strategies (a buy and a sell). At each time step, each participant decides of his optimal control given the current state of the order book. They send their orders at the same time. The control of the HFT is executed first, then this is the turn of the MM, and finally the controls of the {Volume and VWAP trading algorithms} are executed (we choose the order among {them} randomly, according to a uniform distribution). Since the MM and the {Volume and VWAP trading algorithms} decide before seeing the action of the HFT, their controls may not be applied or only partially, depending of the state of the order book after the previous participants have played. 
The exogenous randomness only comes from the simulation of the stock-futures spread process $S$ and from the new queues created when one queue is depleted. If the bid queue is depleted, then the bid price moves down. If the bid-ask spread is equal to two ticks, then the ask price moves down as well. The other way round if the ask queue is depleted. If the bid (resp.~the ask) price moves down (resp.~up), we consider that this is a discovered limit and its size is chosen as in Section \ref{subsec: numerics MM}: $10$ units with probability $60\%$, $5$ units with probability $25\%$ and $12$ units with probability $15\%$.  If the bid (resp.~the ask) price move up (resp.~down), which can happen if the bid-ask spread is already of two ticks, we consider that this is a created limit and its size is chosen again as in Section \ref{subsec: numerics MM}: 
  $2$ units with probability $60\%$, $1$ unit with probability $25\%$ and $3$ units with probability $15\%$. There is no other randomness, the rest of the dynamics is du to the HFT, the MM and the four {other trading algorithms}.  
\vs2

 Note that only the VWAP {trading algorithms} are forced to trade, when they reach the (upper or lower) limit of their prescribed inventory path.
 In principle, the MM plays aggressive orders only when he needs to adjust quickly his inventory. For a typical path of the  stock-futures spread, the HFT also has no incentive to send aggressive orders, except to adjust his inventory. Without trades, the Volume {trading algorithms} do not act as well (because the market volume does not move). Therefore, if the MM and the HFT have a zero initial inventory, we expect to have to wait for the VWAP {trading algorithms} to initiate first {aggressive} orders, and starts the whole dynamics. 
 
 In our illustration, we start with an initial state in which the   HFT  and the   MM   have a zero initial inventory.  Each VWAP {trading algorithm} has to buy/sell 75 stocks within   {$5$} minutes of trading. This corresponds to an average of $0.5$ trades per second, which is consistent with the priors of the MM and the HFT.  
 
\begin{figure}[h!]
 \includegraphics[width=\textwidth,height=0.86\textheight]{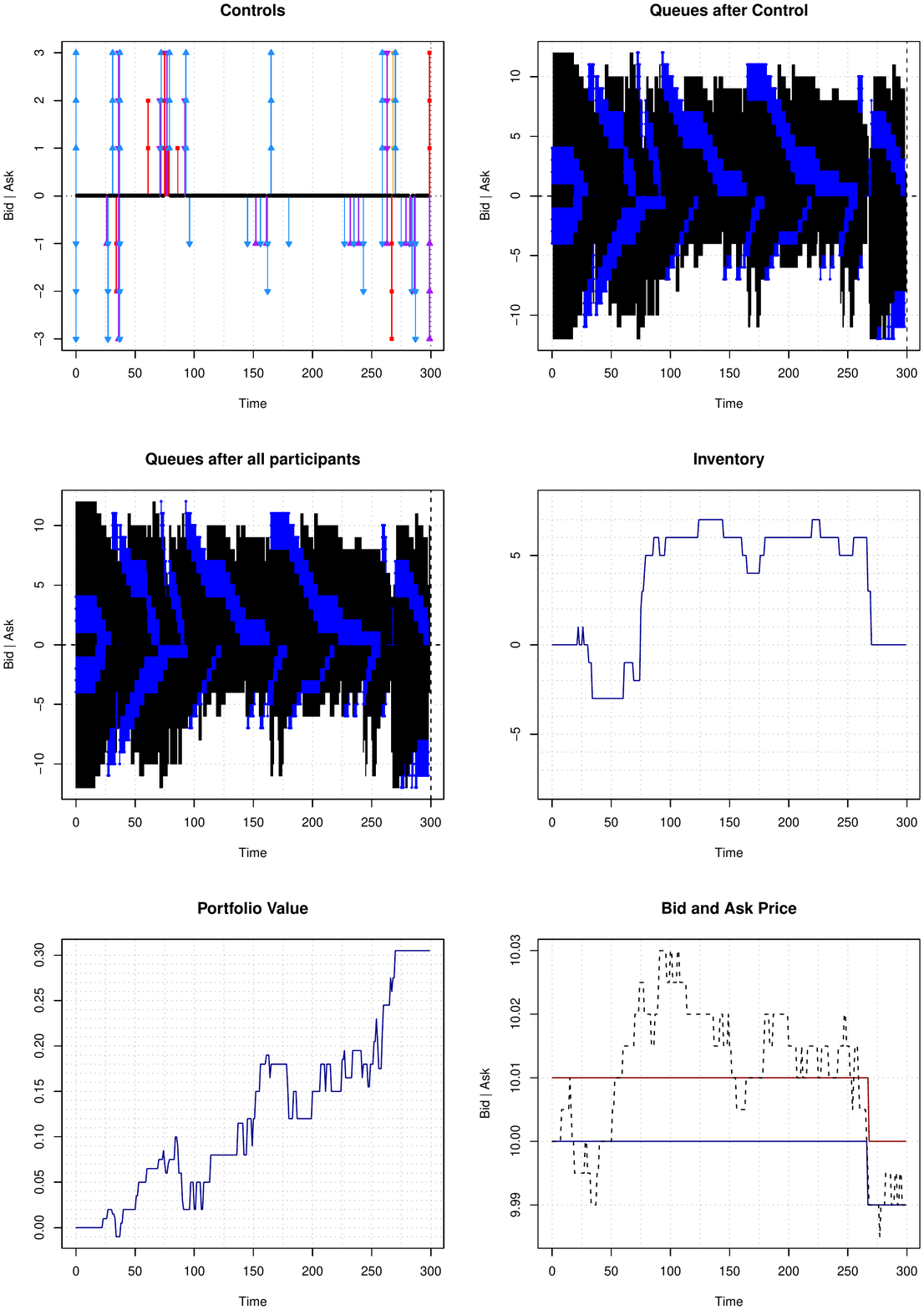}
 \caption{Optimal strategy of the High Frequency Trader when agents play together.} 
\label{fig: together_HFT}
\end{figure}

\begin{figure}[h!]
 \includegraphics[width=\textwidth,height=0.86\textheight]{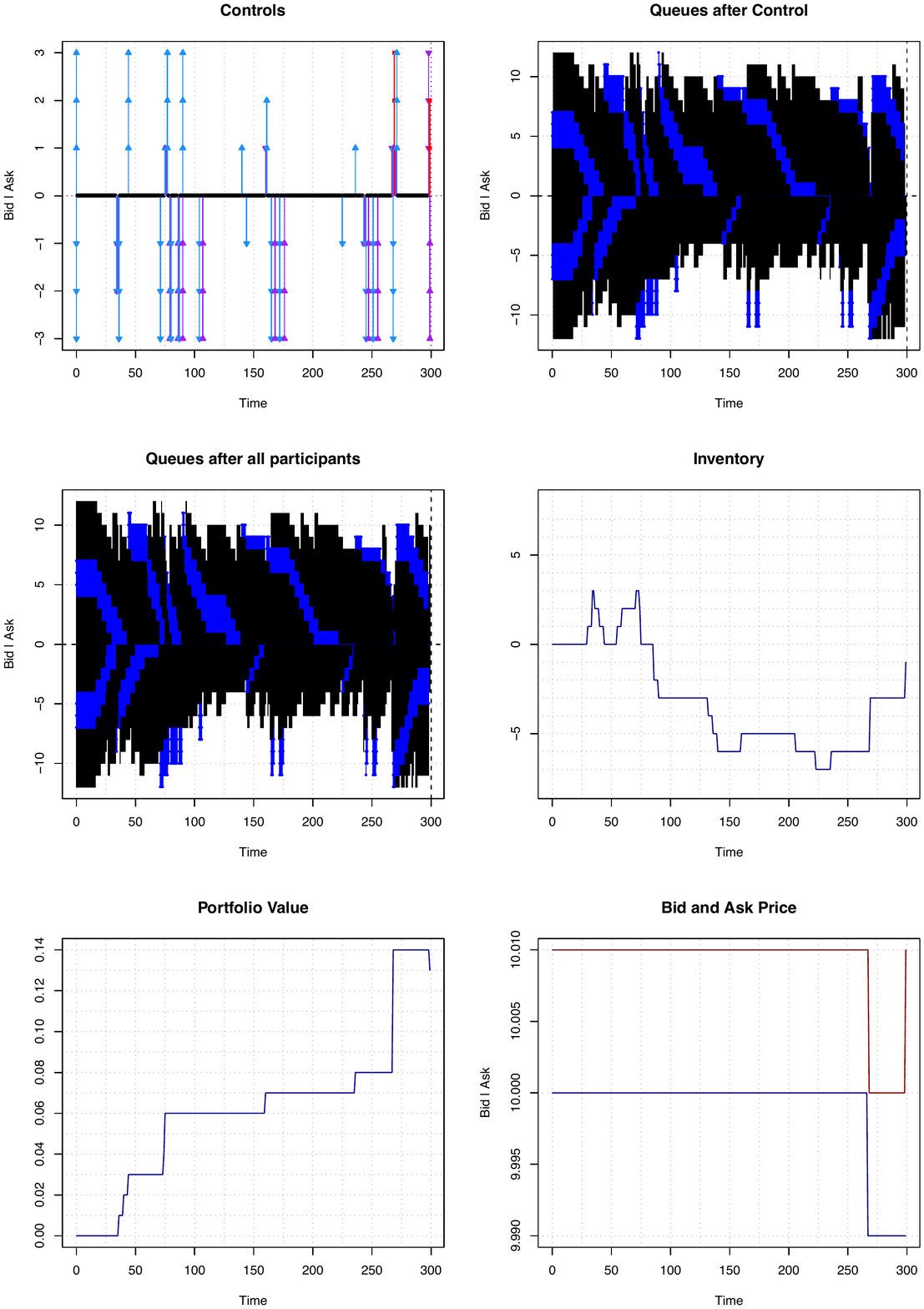}
 \caption{Optimal strategy of the Market Maker  when agents play together.} 
\label{fig: together_MM}
\end{figure}

\begin{figure}[h!]
 \includegraphics[width=\textwidth,height=0.86\textheight]{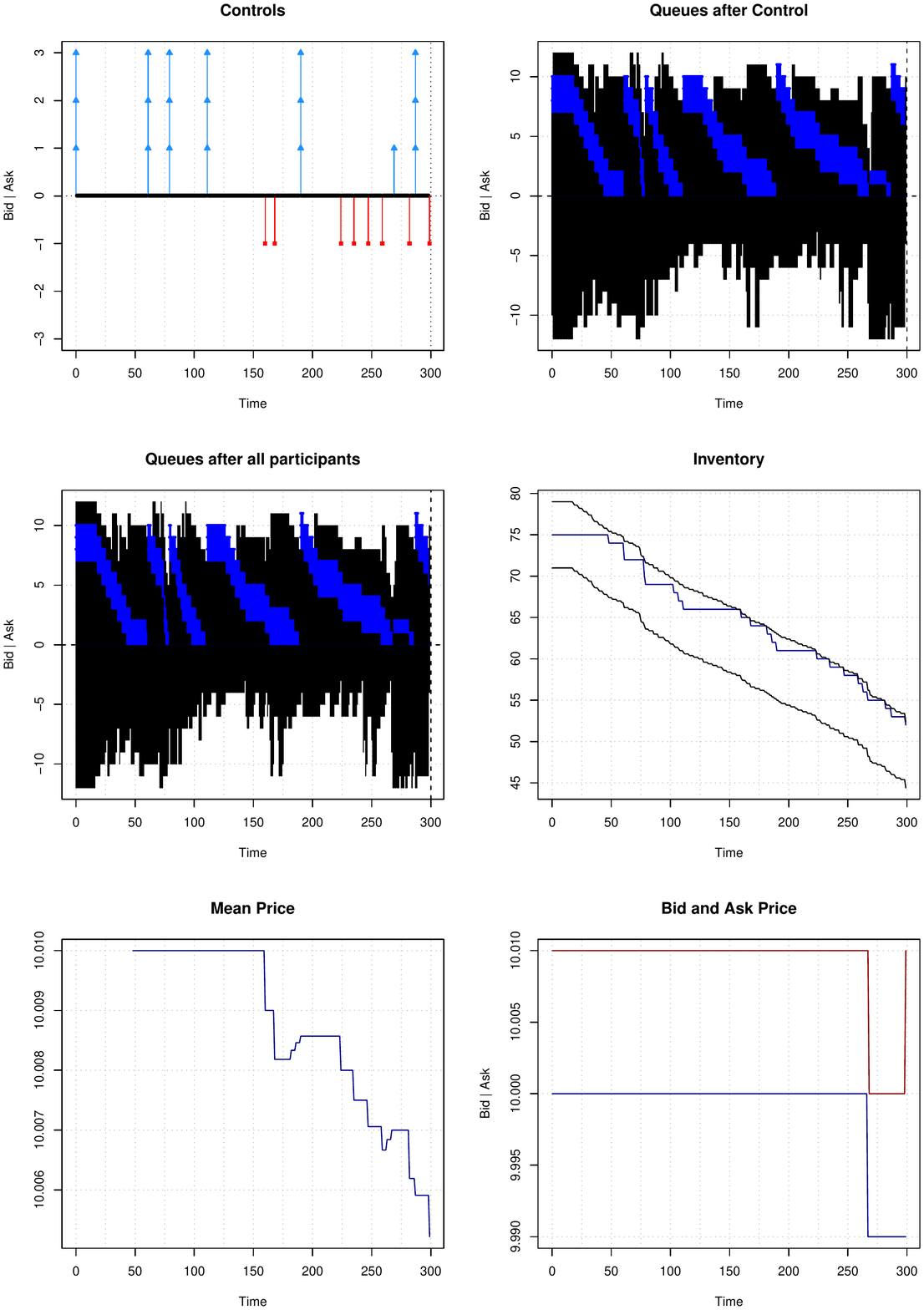}
 \caption{Optimal strategy of the Volume {trading algorithm} (seller)  when agents play together.} 
\label{fig: together_volV}
\end{figure}

\begin{figure}[h!]
 \includegraphics[width=\textwidth,height=0.86\textheight]{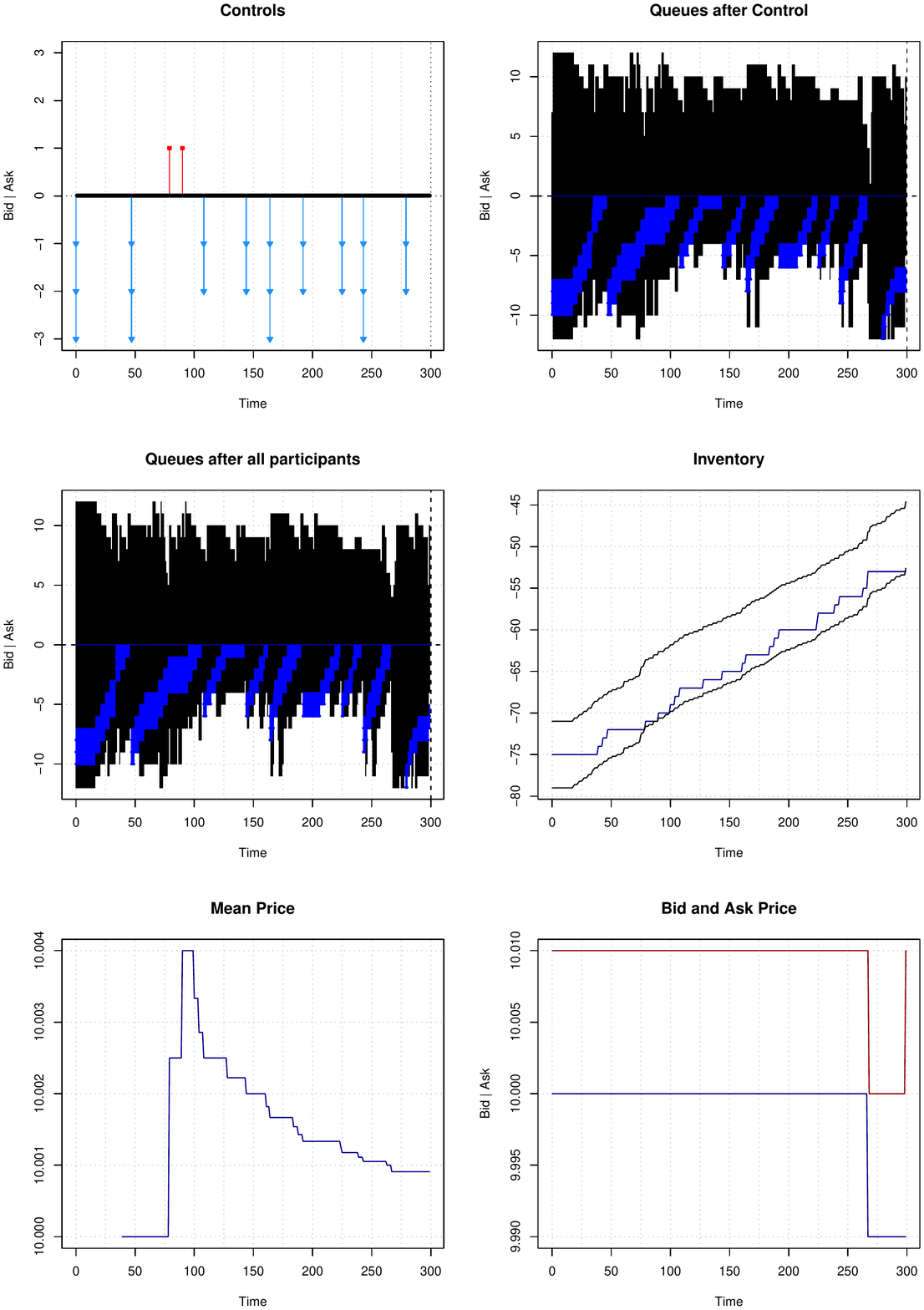}
 \caption{Optimal strategy of the Volume {trading algorithm} (buyer)  when agents play together.} 
\label{fig: together_volA}
\end{figure}

\begin{figure}[h!]
 \includegraphics[width=\textwidth,height=0.86\textheight]{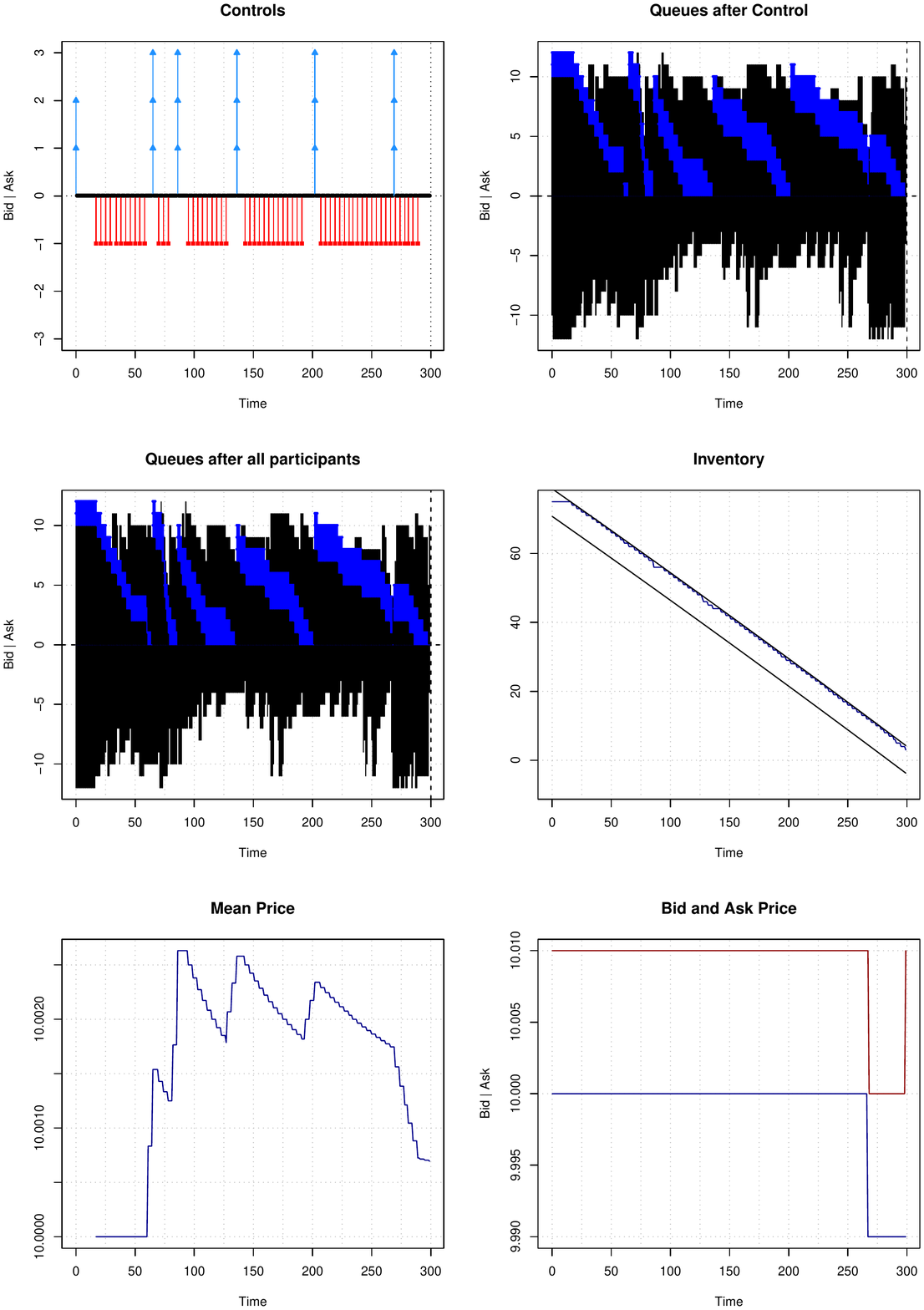}
 \caption{Optimal strategy of the VWAP {trading algorithm} (seller)  when agents play together.} 
\label{fig: together_vwapV}
\end{figure}

\begin{figure}[h!]
 \includegraphics[width=\textwidth,height=0.86\textheight]{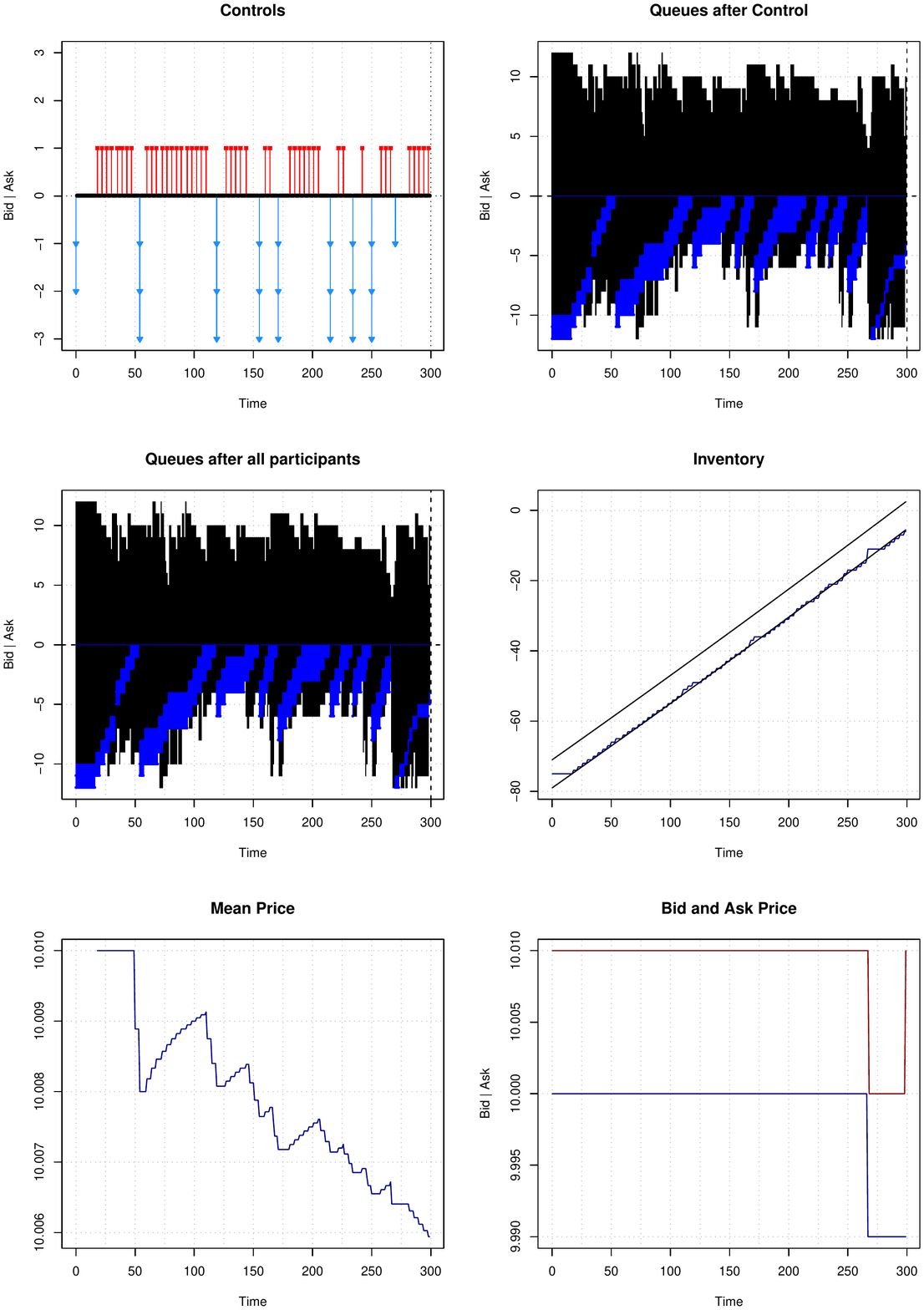}
 \caption{Optimal strategy of the VWAP {trading algorithm} (buyer)  when agents play together.} 
\label{fig: together_vwapA}
\end{figure}

 The Figures \ref{fig: together_HFT}, \ref{fig: together_MM}, \ref{fig: together_volV}, \ref{fig: together_volA}, \ref{fig: together_vwapV} and \ref{fig: together_vwapA} have the same interpretation  as in Section \ref{subsec: numerics MM} except that now the position of the agent in the queue is in blue in the top right and middle left graphics (the black part corresponding to the other participants). The top right graphic is the state of the book just after the control of the  agent is executed, the middle left is the state of the order book after the controls of all  the participants  have been executed. Note surprisingly, the trades are essentially due to the VWAP {trading algorithms}, to which the Volume {trading algorithms} need to adjust. The aggressive orders sent by the HFT and the MM are either cancellations or corrections of their inventory near the terminal time, except around time $35$ at which the HFT clearly wants to take a position on the stock-futures spread because it is very low.  
Note that the first price jump is due to an aggressive sell order of the HFT, slightly before time 275, which pushes the bid price down. He just after sends a limit sell order in the spread, and the ask price moves down as well. Since he is executed first and all players decide at the same time, he does not give to the others the opportunity to take this position in the spread. The last price move, just before $T$, is   due to the HFT and the MM. This time  the HFT and the MM send  aggressive buy orders that deplete the queue (we do not see the price move on the graphics of the HFT, because they provide the prices after the execution of the order of the agent).   As in Section \ref{sec: numeric HFT}, his position follows the stock-futures spread. In particular, between time 50 and 100, one can observe that he waits before sending a new sell limit order because the spread is gone up and he wants to buy stocks to short-sell the futures. When the spread starts going done, he takes a stronger position at the ask than at the bid, to start selling back the stock. As for the MM, he finishes with an inventory equal to $-1$, that will eventually be liquidated at   $10.010$, which explains the downward jump  of his portfolio value at $T$.  When his inventory is very low, he balances between putting limit buy orders and letting the price go down by not supporting the bid limit. It is not successful until the price eventually moves down around time 275.

\section{Conclusion}

We have proposed a simplified but still realistic modeling of an order book, whose dynamics depends on the current imbalance. We have derived optimal strategies for three key actors: Market Makers, High Frequency Traders and Institutional Brokers (volume and VWAP trading algorithms). In this model, optimal strategies can be estimated numerically. Simulations show how complex optimal strategies are when actors believe that the imbalance actually has an impact on the dynamics of order flows. Our numerical simulations show that the construction of a realistic market simulator is feasible. A desirable next step would be to consider realistic proportions of market participants for the simulations proposed in Section \ref{sec : Simulation of the global market} and to see whether the market statistics used to calibrate the different strategies can be retrieved through the simulated interactions of the various players. If this is the case, one can imagine trying to exhibit different market patterns depending on the nature and proportions of actors currently acting.

\vspace{2mm}
\noindent {\bf Acknowledgment:} This research is part of a Cemracs 2017 project and benefited from the support of the Initiative de Recherche from  Kepler-Chevreux and Coll\`{e}ge de France. We are very grateful for the support of P.~Besson and his team.

\newpage
 
\newpage
 
\newpage
 
\newpage
 
\newpage
 
\newpage
 
\bibliographystyle{plain}

\end{document}